\documentclass[technote,draftcls,onecolumn,11pt,twoside]{IEEEtran}

\usepackage[utf8]{inputenc}
\usepackage{cite}

\usepackage{latexsym,epsfig,amssymb,amsfonts,eucal,graphicx}
\usepackage{subfigure}
\usepackage[cmex10]{amsmath}
\usepackage{slantsc}
\usepackage{enumerate}
\usepackage{setspace}
\usepackage{graphicx}
\usepackage{epstopdf}
\usepackage{mdwmath}
\usepackage{mdwtab}
\usepackage{color}
\usepackage{soul}
\usepackage{multirow}
\usepackage{array}
\usepackage{booktabs}

\newcommand{\markchanges}{black}

\begin{document}
\title{Spatial Multiplexing of QPSK Signals with a Single Radio:\\ Antenna Design and Over-the-Air Experiments}

\author{Mohsen~Yousefbeiki, %\IEEEmembership{Member,~IEEE,}
		Andrew~C.~M.~Austin, %\IEEEmembership{Member,~IEEE,}
        Juan~R.~Mosig, 		
		Andreas~Burg, %\IEEEmembership{Member,~IEEE,}        
        and Julien~Perruisseau-Carrier% <-this % stops a space
\thanks{This work was supported by the Swiss National Science Foundation (SNSF) under Grant No. 133583 and the Korean Electronics and Telecommunications Research Institute (ETRI).}% <-this % stops a space
\thanks{M. Yousefbeiki was with the Laboratory of Electromagnetics and Antennas (LEMA), Ecole Polytechnique F\'{e}d\'{e}rale de Lausanne (EPFL), 1015 Lausanne, Switzerland (e-mail: mohsen.yousefbeiki@alumni.epfl.ch).}% <-this % stops a space
\thanks{J. R. Mosig is with the Laboratory of Electromagnetics and Antennas (LEMA), Ecole Polytechnique F\'{e}d\'{e}rale de Lausanne (EPFL), 1015 Lausanne, Switzerland.}% (e-mail: mohsen.yousefbeiki@epfl.ch).}% <-this % stops a space
\thanks{A.C.M. Austin and A. Burg are with the Telecommunications Circuits Laboratory (TCL), Ecole Polytechnique F\'{e}d\'{e}rale de Lausanne (EPFL), 1015 Lausanne, Switzerland.}
\thanks{J. Perruisseau-Carrier (deceased) was with the group for Adaptive MicroNanoWave Systems, Ecole Polytechnique F\'{e}d\'{e}rale de Lausanne (EPFL), 1015 Lausanne, Switzerland.}}


\markboth{\textit{submitted to} IEEE Transactions on Antennas and Propagation}%
{Shell \MakeLowercase{\textit{et al.}}: Bare Demo of IEEEtran.cls for Journals}



\maketitle

\begin{abstract}
The paper describes the implementation and performance analysis of the first fully-operational beam-space MIMO antenna for the spatial multiplexing of two QPSK streams. The antenna is composed of a planar three-port radiator with two varactor diodes terminating the passive ports. Pattern reconfiguration is used to encode the MIMO information onto orthogonal virtual basis patterns in the far-field.
%The antenna functionality is demonstrated by excellent agreement between the measured and simulated results as well as by successful over-the-air MIMO transmission-reception of two QPSK signals from a single RF source. 
A measurement campaign was conducted to compare the performance of the beam-space MIMO system with a conventional $2\!\times\!2$ MIMO system under realistic propagation conditions. Propagation measurements were conducted for both systems and the mutual information and symbol error rates were estimated from Monte-Carlo simulations over the measured channel matrices. The results show the beam-space MIMO system and the conventional MIMO system exhibit similar finite-constellation capacity and error performance in NLOS scenarios when there is sufficient scattering in the channel. In comparison, in LOS channels, the capacity performance is observed to depend on the relative polarization of the receiving antennas.

%the capacity performance of beam-space MIMO observed in the considered LOS channel depends on the relative polarization of the receiving antennas, which is expected owing to the fact that the basis patterns are not omni-directional and similarly polarized.

\end{abstract}


% \fi
\IEEEpeerreviewmaketitle

\section{Introduction}
%\IEEEPARstart{S}{patial} 

\textcolor{\markchanges}{Beam-space multiple-input multiple-output (MIMO) systems have recently emerged as an efficient alternative to conventional MIMO systems. In beam-space MIMO, multiple RF chains and antennas at the transmitter are replaced with a more economical single-radio and a pattern-reconfigurable antenna, while a conventional MIMO architecture is maintained at the receiver~\cite{kalis2008}. The beam-space MIMO transmit antenna exactly emulates conventional MIMO transmission at each symbol interval, by mapping multiple symbols onto an orthogonal set of virtual basis patterns in the far-field. By only requiring a single RF chain, beam-space MIMO provides a way to potentially exploit MIMO benefits in low-end user terminals or sensor nodes, with strict size and complexity constraints.}

%One of the major applications of beam-space MIMO is 

%Beam-space MIMO paves the way for exploiting MIMO benefits, such as spectral efficiency enhancement in low-end user terminals, and sensor nodes with strict size and complexity constraints, 

%However, the actual design of antenna systems for beam-space MIMO requires addressing various practical aspects that are commonly overlooked in theoretical and conceptual developments. 

\textcolor{\markchanges}{One of the primary challenges of beam-space MIMO is the design of low-complexity reconfigurable antenna systems capable of creating the radiation patterns necessary to support higher order modulation schemes, such as phase shift keying (PSK) and quadrature amplitude modulation (QAM).} As radiation pattern reconfiguration is typically achieved by changing the impedance of variable loads embedded in the antenna structure, e.g.,~\cite{julien_eumc,alrabadi_tap,my_tap14}, successful data multiplexing in beam-space MIMO requires finding and realizing the correct impedance values. These impedance values not only depend on the symbols to be multiplexed, but are also very dependent on how the basis patterns for the pattern decomposition are defined.  Moreover, the antenna system performance, particularly in terms of impedance matching, will typically vary as the impedance of the variable loads is changed. \textcolor{\markchanges}{For the realization of such load the use of passive-only components is preferred, as the complexity and stability concerns of active loads offset the benefits. The most advanced beam-space MIMO antenna demonstrated today only supports BPSK modulation~\cite{alrabadi_tap,my_tap14}.}

\textcolor{\markchanges}{The performance of MIMO systems in real environments depends on the availability of multiple independent channels~\cite{Oestges}. The orthogonality of the basis patterns in beam-space MIMO---similar to the orthogonality of the radiation patterns in conventional MIMO---leads to uncorrelated channels, provided there are a sufficient number of randomly and uniformly located scatterers to form a uniform scattering medium in the entire space~\cite{Flaviis}. In this case the symbol streams transmitted from a beam-space MIMO antenna experience multipath fading similar to conventional MIMO transmission. Therefore, in rich-scattering environments beam-space MIMO can match the performance of conventional MIMO with a corresponding constellation alphabet. However, in real propagation environments the angular spread of the multipath components is often far from ideal, and thus modulating orthogonal radiation (or basis) patterns does not guarantee the decorrelation of the MIMO channels. In the case of conventional MIMO, numerous studies have been conducted to investigate the system performance in real channels. For example, in~\cite{rich_scattering,real_channel1,real_channel2} it was shown that inadequate scattering leads to channel decorrelation degradation, and this has a significant impact on the performance limits of MIMO systems. However, little is known about the performance of beam-space MIMO in real propagation environments. Unlike conventional MIMO, where transmit antennas typically have omni-directional similarly-polarized radiation patterns, basis patterns in beam-space MIMO have different polarization characteristics and propagate from a single radiating structure towards different directions in space. Accordingly, the performance of beam-space MIMO may be more affected by the channel geometry. }

\textcolor{\markchanges}{\emph{Contributions}: 
To remove the BPSK limitation of beam-space MIMO antennas, the authors in~\cite{my_tcomm} proposed a strategy to multiplex PSK data streams (of any modulation order), using only purely reactive reconfigurable loads, while ensuring a state-independent impedance matching at the antenna input. In this paper, the strategy proposed in~\cite{my_tcomm} is extended to design and implement the first fully-operational beam-space MIMO antenna capable of transmitting two QPSK signals. The design procedure is described and the results demonstrating the efficiency of the approach are presented. The functionality of the proposed beam-space MIMO antenna is then confirmed through over-the-air experiments using a testbed constructed from off-the-shelf hardware and software components. A measurement campaign is conducted to compare the performance of beam-space MIMO with conventional MIMO in real propagation environments. }

\textcolor{\markchanges}{\emph{Outline}: 
The paper is organized as follows. Section~II presents the design, fabrication, and measurements of the proposed beam-space MIMO antenna. The beam-space MIMO testbed and validation results are presented in Section~III. In Section~IV, the experimental procedure and performance criteria are discussed. The results obtained from the over-the-air experiments are presented and analyzed in Section~V. The paper is breifly summarized in section VI.}

%%%%%%%%%%%%%%%%%%%%%%%%%%%%%%%%%%%%%%%%%%%%%%%%%%%%%%%%%%%%%%%%%%%%%%%%%%%%%%%%%
%%%%%%%%%%%%%%%%%%%%%%%%%%%%%%%%%%%%%%%%%%%%%%%%%%%%%%%%%%%%%%%%%%%%%%%%%%%%%%%%%%%%
%%%%%%%%%%%%%%%%%%%%%%%%%%%%%%%%%%%%%%%%%%%%%%%%%%%%%%%%%%%%%%%%%%%%%%%%%%%%%%%%%%%%%%
\section{Beam-Space MIMO Antenna for QPSK Signaling}   \label{secII}
\subsection{Theory and Multiplexing Strategy} \label{secIIA}
\begin{figure}%%[!t]
\centering
\includegraphics[width=3.0in]{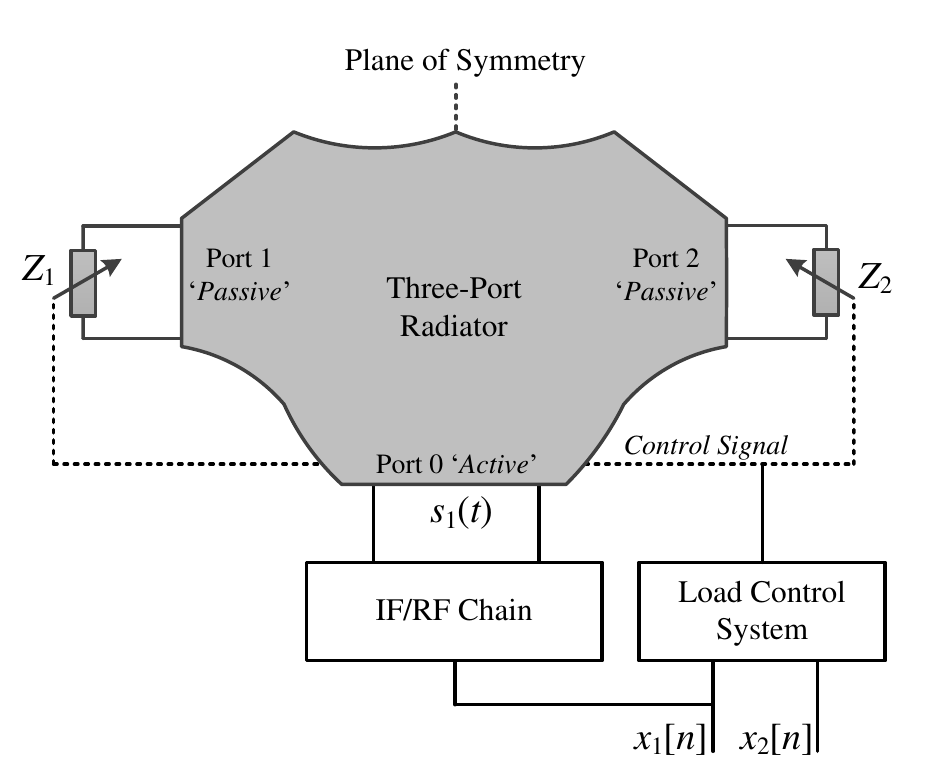}
\caption{Schematic of the beam-space MIMO antenna system reported in~\cite{my_tcomm}. The three-port radiator is fully described by a 3-by-3 scattering matrix, $\mathbf{S}$, and three embedded radiation patterns.}
\label{fig:fig1}
\end{figure}

\textcolor{\markchanges}{The key idea in beam-space MIMO is to engineer the instantaneous radiation pattern of a single-feed reconfigurable antenna for each symbol interval, such that multiple independent data streams directly modulate a predefined set of orthogonal basis patterns in the far-field.} Fig.~\ref{fig:fig1} shows a symbolic representation of the antenna system proposed in~\cite{my_tcomm} for efficient and low-complexity beam-space MIMO transmission of PSK signals. The antenna is composed of a symmetric three-port radiator and two variable loads $Z_1$ and $Z_2$ connected to the passive ports. The first symbol stream $x_1$ is up-converted and fed into the antenna active port. To multiplex each arbitrary symbol pair of $\{x_1, x_2\}$ with a combination ratio of 
%%%%%%%%%%%%%%%%%%%%%%%%%%%%%
\begin{equation}\label{eq:xr}
\bar{x} = {{{x_2}} \mathord{\left/
 {\vphantom {{{x_2}} {{x_1}}}} \right.
 \kern-\nulldelimiterspace} {{x_1}}} \,,
\end{equation}
%%%%%%%%%%%%%%%%%%%%%%%%%%%%%
a \emph{load control system} reconfigures the variable loads, so that the antenna instantaneous radiated field becomes
%%%%%%%%%%%%%%%%%%%%%%%%%%%%%
\begin{equation}\label{eq:bsmimo_concept}
\boldsymbol{\mathcal{E}}_{\rm{inst}}^{\{ {x_1},{x_2}\} }(\theta ,\varphi ) = {x_1}\boldsymbol{\mathcal{B}}_1(\theta ,\varphi ) + {x_2}\boldsymbol{\mathcal{B}}_2(\theta ,\varphi ) \,,
\end{equation}
%%%%%%%%%%%%%%%%%%%%%%%%%%%%%
where $\boldsymbol{\mathcal{B}}_1(\theta ,\varphi )$ and $\boldsymbol{\mathcal{B}}_2(\theta ,\varphi )$ are the basis patterns (as will be defined later in this section). Defining $\boldsymbol{\mathcal{E}}_{\rm{unit}}(\theta,\varphi)$ as the instantaneous radiated field for a unit power excitation, (\ref{eq:bsmimo_concept}) can be simplified as
%%%%%%%%%%%%%%%%%%%%%%%%%%%%%
\begin{equation}\label{eq:bsmimo_concept2}
\boldsymbol{\mathcal{E}}_{\rm{unit}}^{\{ \bar{x}\} }(\theta ,\varphi ) = \boldsymbol{\mathcal{B}}_1(\theta ,\varphi ) + \bar{x}\boldsymbol{\mathcal{B}}_2(\theta ,\varphi ) \,,
\end{equation}
%%%%%%%%%%%%%%%%%%%%%%%%%%%%%
showing the dependency of the multiplexing relation on the ratio of the two symbols to be multiplexed, and not to the specific values, i.e., $\{x_1, x_2\}$. Therefore, transmitting any symbol pair with the same ratio, $\bar{x}$, the variable loads $Z_1$ and $Z_2$ are reconfigured to the same impedance values, i.e., $Z_1^{\{ \bar{x}\} }$ and $Z_2^{\{ \bar{x}\} }$, respectively. 

%%%%%%%%%%%%%%%%%%%%%%%%%%%%%%%%%%%%%%
%%%%%%%%%%%%%%%%%%%%%%%%%%%%%%%%%%%%%%%%%%

According to (\ref{eq:bsmimo_concept2}), the two basis patterns can be written as
%%%%%%%%%%%%%%%%%%%%%%%%%%%%%
\begin{subequations}\label{eq:basis}
\begin{align}
&\boldsymbol{\mathcal{B}}_1(\theta ,\varphi ) = \frac{\boldsymbol{\mathcal{E}}_{\rm{unit}}^{\{ +1\} }(\theta ,\varphi )+\boldsymbol{\mathcal{E}}_{\rm{unit}}^{\{ -1\} }(\theta ,\varphi )}{2}  \\
&\boldsymbol{\mathcal{B}}_2(\theta ,\varphi ) = \frac{\boldsymbol{\mathcal{E}}_{\rm{unit}}^{\{ +1\} }(\theta ,\varphi )-\boldsymbol{\mathcal{E}}_{\rm{unit}}^{\{ -1\} }(\theta ,\varphi )}{2} .
\end{align}
\end{subequations}
%%%%%%%%%%%%%%%%%%%%%%%%%%%%%
It has been proven in~\cite{my_tcomm} that the orthogonality of the basis patterns is guaranteed for any arbitrary choice of $Z_1^{\{  + 1\} } = Z_2^{\{  - 1\} }$ and $Z_2^{\{  + 1\} } = Z_1^{\{  - 1\} }$. However, applying the constraint of only purely reactive loads (under PSK signaling) limits the freedom in the selection of either $Z_1^{\{  - 1\} } = jX_1^{\{  - 1\} }$, or $Z_2^{\{  - 1\} } = jX_2^{\{  - 1\} }$~\cite{my_tcomm}. Taking $X_1^{\{-1\} }$ as the free parameter, the required reactance values can be obtained from
%%%%%%%%%%%%%%%%%%%%%%%%%%%%%
\begin{equation}\label{eq:reactances}
X_1^{\{\bar{x} \}} = X_2^{\{-\bar{x} \}} =  - {Z_0}\frac{{{c_1}X_1^{\{ -1\}} + {c_2}}}{{{d_1}X_1^{\{ -1 \}} + {d_2}}} \,,
\end{equation}
%%%%%%%%%%%%%%%%%%%%%%%%%%%%%
where $Z_0$ is the reference impedance for the scattering matrix, and for compactness we denote
%%%%%%%%%%%%%%%%%%%%%%%%%%%%%
\begin{align}
&{c_1} = 2{\mathop{\rm Im}\nolimits} \left\{ \Delta  \right\}\cos \dfrac{\arg (\bar{x})}{2} + (1 - {\left| \Delta  \right|^2})\sin \dfrac{\arg (\bar{x})}{2},\nonumber\\
&{c_2} = {Z_0}{\left| {1 + \Delta } \right|^2}\cos \dfrac{\arg (\bar{x})}{2},\nonumber\\
&{d_1} = {\left| {1 - \Delta } \right|^2}\cos \dfrac{\arg (\bar{x})}{2},\nonumber\\
&{d_2} = {2{Z_0}{\mathop{\rm Im}\nolimits} \left\{ \Delta  \right\}\cos \dfrac{\arg (\bar{x})}{2} - {Z_0}(1 - {{\left| \Delta  \right|}^2})\sin \dfrac{\arg (\bar{x})}{2}} ,\nonumber\\
&\Delta  = {S_{11}} - {S_{21}}.\nonumber
\end{align}
%%%%%%%%%%%%%%%%%%%%%%%%%%%%%
Since PSK modulation provides rotationally symmetric constellations, (\ref{eq:reactances}) implies exactly the same set of reactance values are required at both passive ports.

The multiplexing strategy outlined above ensures the reflection coefficient at the antenna active port, $\Gamma_{\rm{tot}}$, (given by (45) in~\cite{my_tcomm}) is a function of only the radiator S-matrix, and therefore remains constant regardless of the symbol combination ratio, $\bar{x}$, and the selection of $X_1^{\{-1\} }$. Similarly, the power, $\mathcal{P}_{\mathcal{B}_1}$ and $\mathcal{P}_{\mathcal{B}_2}$, radiated by the basis patterns, $\boldsymbol{\mathcal{B}}_1(\theta ,\varphi )$ and $\boldsymbol{\mathcal{B}}_2(\theta ,\varphi )$, (given by (55) in~\cite{my_tcomm}) can be expressed independently of $X_1^{\{-1\} }$ in terms of the S-matrix and the embedded radiation patterns of the radiator. This implies the optimization of the beam-space MIMO antenna system---according to a specific criterion in terms of the basis power imbalance ratio and the input impedance---can be performed on the parameters of the radiator.

\subsection{Design Technique}
\begin{figure}[!t]
\centering
\subfigure[]{\includegraphics[scale=0.42]{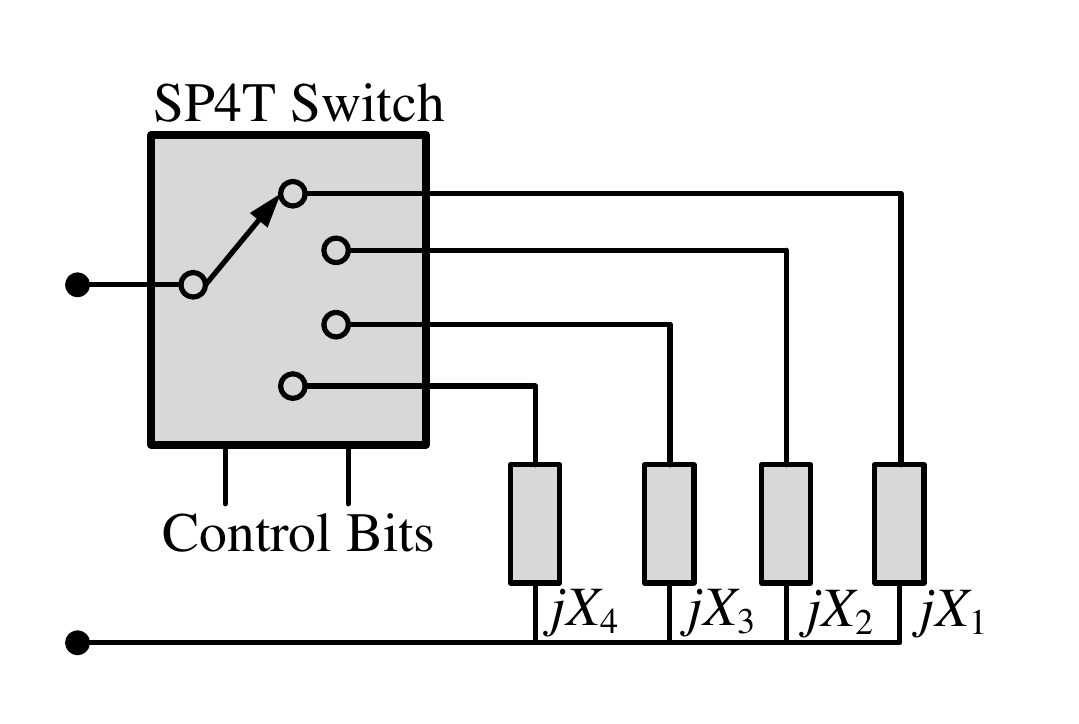} \label{fig:loads1}}
\subfigure[]{\includegraphics[scale=0.42]{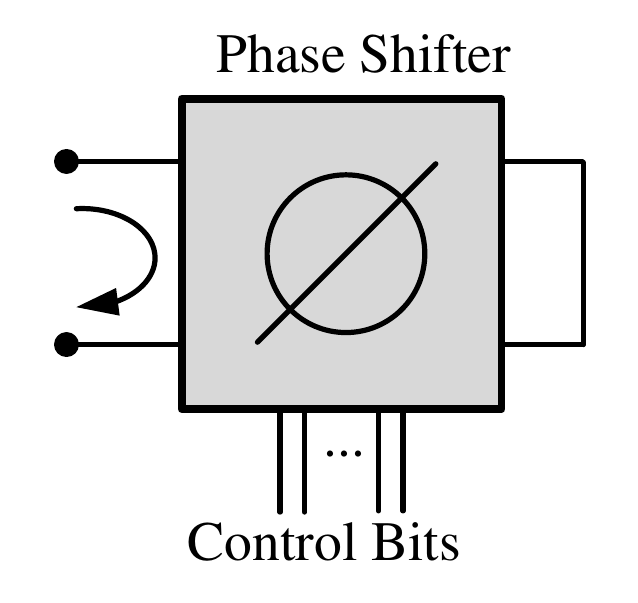} \label{fig:loads2}}
\subfigure[]{\includegraphics[scale=0.42]{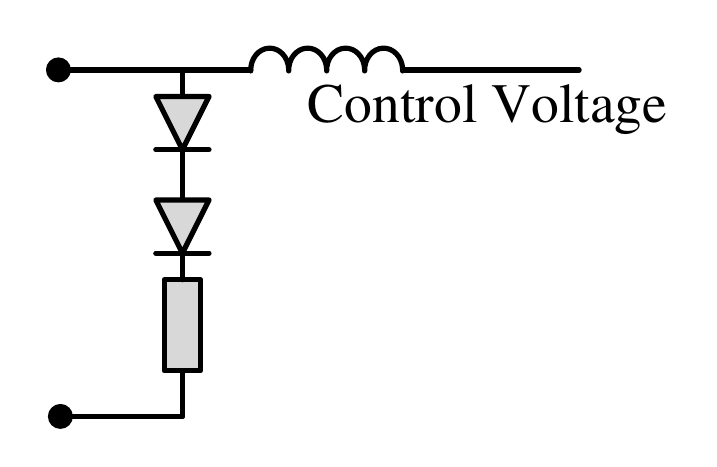} \label{fig:loads3}}
\caption{Various techniques for implementing variable loads of a beam-space MIMO antenna. }
\label{fig:loads_techniques}
\end{figure}

%While designing a beam-space MIMO antenna consisting of a three-port radiator and two variable loads, particular design constraints and performance requirements need to be considered. In particular, 
Successful data multiplexing in beam-space MIMO is very dependent on the hardware technology used for the dynamic reconfiguration. There are several techniques available for realizing the variable loads, with different performance specifications in terms of switching speed, controllability, sensitivity, linearity, and reactance tuning range. As a first approach, microwave switches can be employed to switch among different load circuits, each providing the required reactance for a given symbol combination ratio. For instance, in the case of QPSK modulation where there are four distinct values of $\bar{x}$ (i.e., $\bar{x}=\{\pm1,\pm j\}$), each variable load can be realized using a SP4T \textcolor{\markchanges}{(single-pole four-throw)} switch---or an array of three SPDT \textcolor{\markchanges}{(single-pole double throw)} switches---and four load circuits, as depicted in Fig.~\ref{fig:loads1}. The proper load circuit can be selected (according to the current symbol combination ratio) using a 2-bit control signal.

%%%%%%%%%%%%%%%%%%%%%%%%%%%%
%%%%%%%%%%%%%%%%%%%%%%%%%%%%%%%%%%

Variable reactive loads may also be realized by controlling the phase of the signal reflected from a short-circuited high-precision phase shifter, as shown in Fig.~\ref{fig:loads2}. However, high-precision phase shifters often suffer from high insertion loss, and are thus not the best option for beam-space MIMO applications.

A third approach consists of using one or multiple p-i-n or varactor diodes directly as variable loads by adjusting the DC bias voltage. In this case, lumped components, such as capacitors and/or inductors, can be used in series or parallel with the diodes, as shown in Fig.~\ref{fig:loads3}, to provide the desired range of reactance tunability. When compared to switch-based load implementation techniques, diode-based variable loads provide a more controllable and finer tuning range, and furthermore require fewer lumped elements, resulting in a more compact realization.

Varactor diodes generally yield close to linear reactance-voltage characteristics, and are thus superior to p-i-n diodes, particularly for shaping the transient response of the variable loads required to control out-of-band radiation~\cite{MohsenThesis}. Since varactor diodes operated in the reverse-bias region have no minority carrier charge storage, the rate at which the depletion region changes its width is fast, potentially offering higher switching speed than p-i-n diodes.

\begin{figure}[!t]
\centering
\includegraphics[width=3.5in]{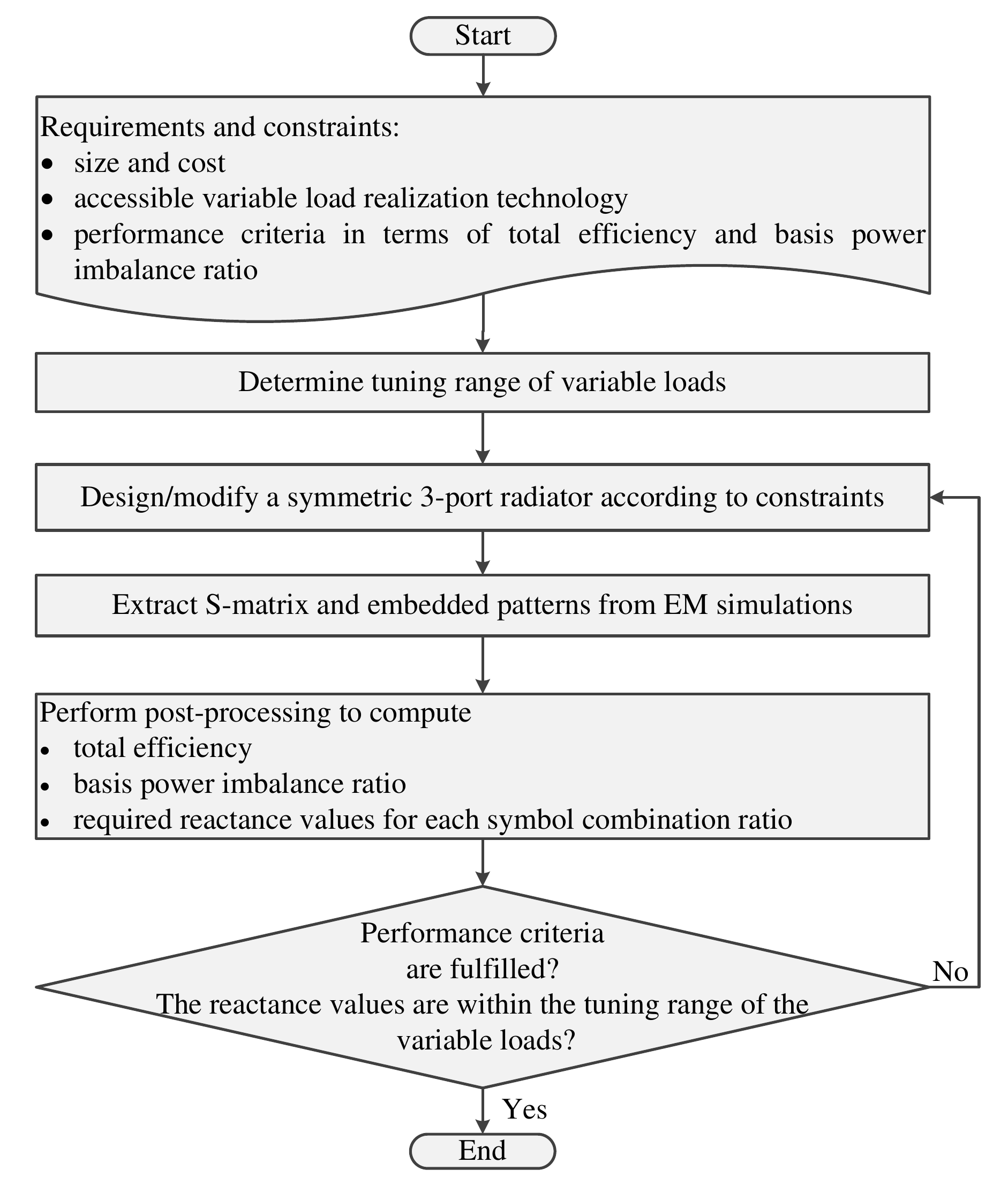}
\caption{Flowchart of the beam-space MIMO antenna design.}
\label{fig:flowchart}
\end{figure}

Once the approach for realizing the variable loads is selected (and the possible tuning range of reactance is determined), the design of the three-port radiator may start. While the radiator must satisfy targeted size and cost limitations, the beam-space MIMO approach imposes another geometrical constraint, namely mirror-image symmetry of the radiator. The three-port radiator, whether in a physically-separate array configuration or an integrated form, can be fully represented by a 3-by-3 S-matrix, and the three angular embedded radiation patterns. This information is used to compute the required reactance values of the variable loads, and to evaluate the beam-space MIMO antenna performance in terms of impedance matching, radiation efficiency, and basis power imbalance ratio $r = \mathcal{P}_{\mathcal{B}_1} \mathord{\left/
 {\vphantom {\mathcal{P}_{\mathcal{B}_1} \mathcal{P}_{\mathcal{B}_2}}} \right.
 \kern-\nulldelimiterspace} {\mathcal{P}_{\mathcal{B}_2}}$. To meet these specifications, the radiator is designed using 
%The radiator must be engineered in such a way that all of the above-mentioned characteristics lie within acceptable limits. This requires 
extensive full-wave simulation-based optimization to determine the placement of the passive ports and the dimensions of the conductor traces. Fig.~\ref{fig:flowchart} depicts a flowchart in which the beam-space MIMO antenna design procedure is described step by step.

%%%%%%%%%%%%%%%%%%%%%%%%%%%%%%%%%%%%%%
%%%%%%%%%%%%%%%%%%%%%%%%%%%%%%%%%%%%%%%%%%

\subsection{QPSK Beam-Space MIMO Antenna Prototype Design}

In this section, we present the design procedure of a compact planar reconfigurable antenna prototype for the beam-space multiplexing of two QPSK signals at a design frequency of 2.45~GHz. A good impedance match at the single active port (i.e., $\left| \Gamma_{\rm{tot}} \right|^2 \le -10$~dB) and nearly balanced basis patterns (i.e., $\left| r \right| \le  3$~dB) were the design requirements to ensure optimum open-loop MIMO performance in rich-scattering environments.   

\begin{figure}[!t]
\centering
\subfigure[]{\includegraphics[width=3.1in]{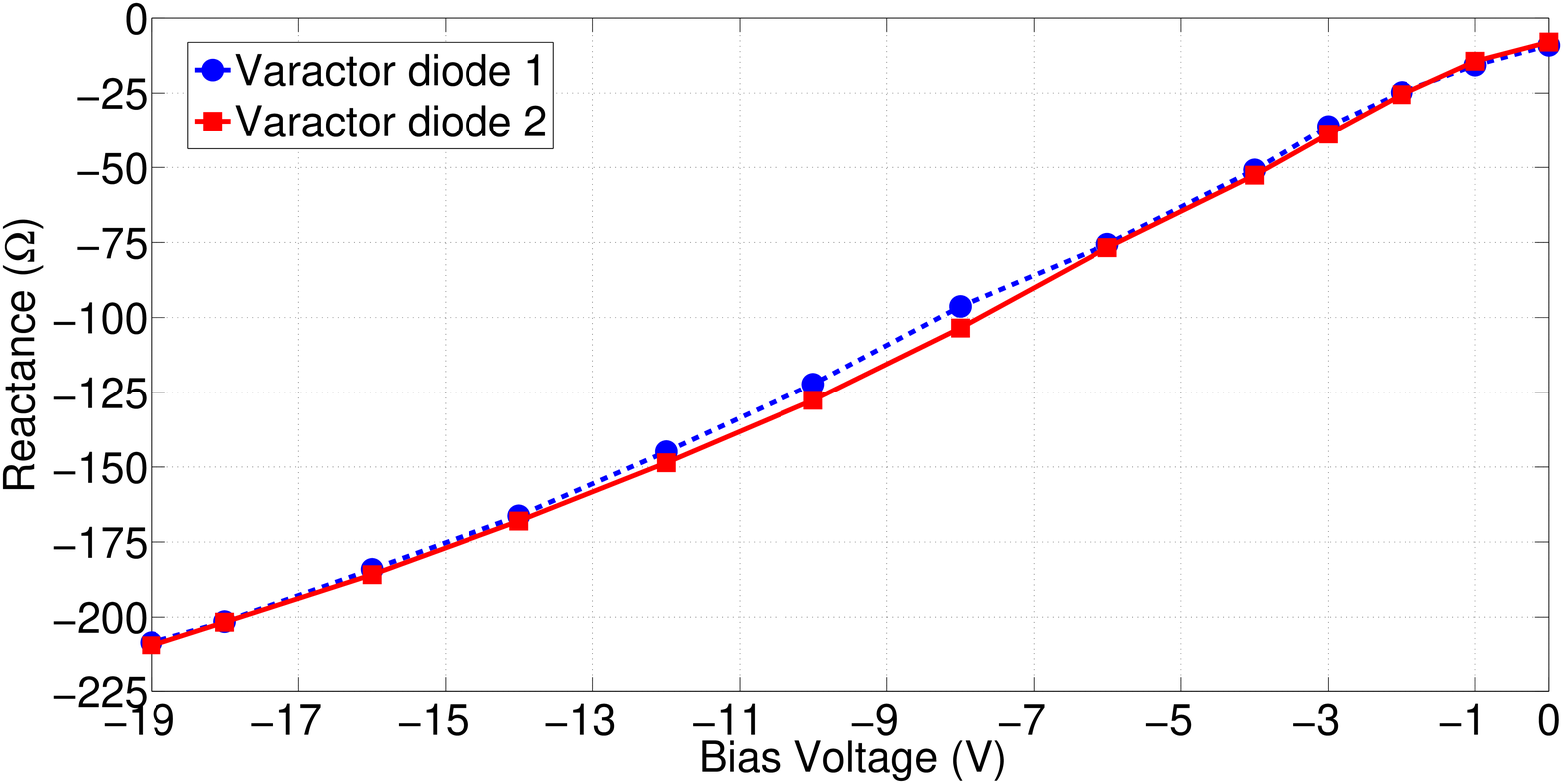} \label{fig:X_AB_vs_V}}
\subfigure[]{\includegraphics[width=3.1in]{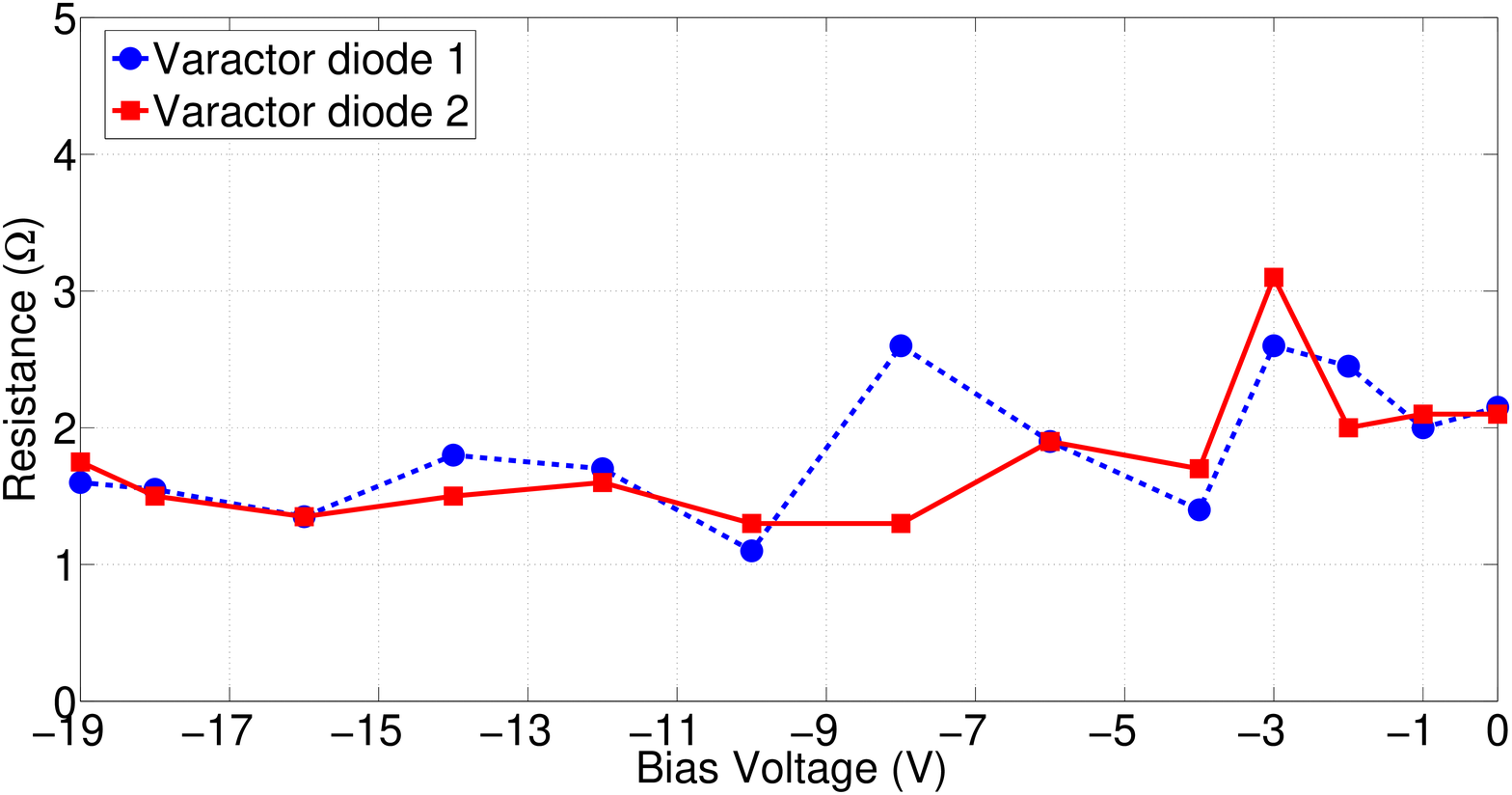} \label{fig:R_AB_vs_V}}
\caption{(a) Voltage-reactance and (b) voltage-resistance corrected curves of the varactor diodes. The reverse breakdown voltage of the diodes is reported to be around $22$~V.}
\label{fig:RX_AB_vs_V}
\end{figure}

Since a smooth voltage-reactance characteristic, high-speed switching and compactness were of our interest in this work, we selected two GaAs hyperabrupt varactor diodes (Aeroflex/Metelics MGV125-22) as the sole mounted components at the passive ports of the radiator. The precise voltage-impedance curves of the varactor diodes were extracted from separate measurements using a thru-reflect-line (TRL) calibration method. The measured impedance curves were corrected for use in the simulations using the technique described in~\cite{my_apmag}. As shown in Fig.~\ref{fig:RX_AB_vs_V}, at the design frequency the two varactor diodes offer a reactance tuning range of [$-9$,$-209$]~$\Omega$ and [$-8$,$-210$]~$\Omega$, respectively, while varying the bias voltage between $0$~V and $-19$~V. The real part of the impedance is negligible (about $2$~$\Omega$), and remains relatively constant over the bias voltage range. 

%%%%%%%%%%%%%%%%%%%%%%%%%%%%
%%%%%%%%%%%%%%%%%%%%%%%%%%%%%%%%%%

\begin{figure}[!t]
\centering
\includegraphics[width=2.1in]{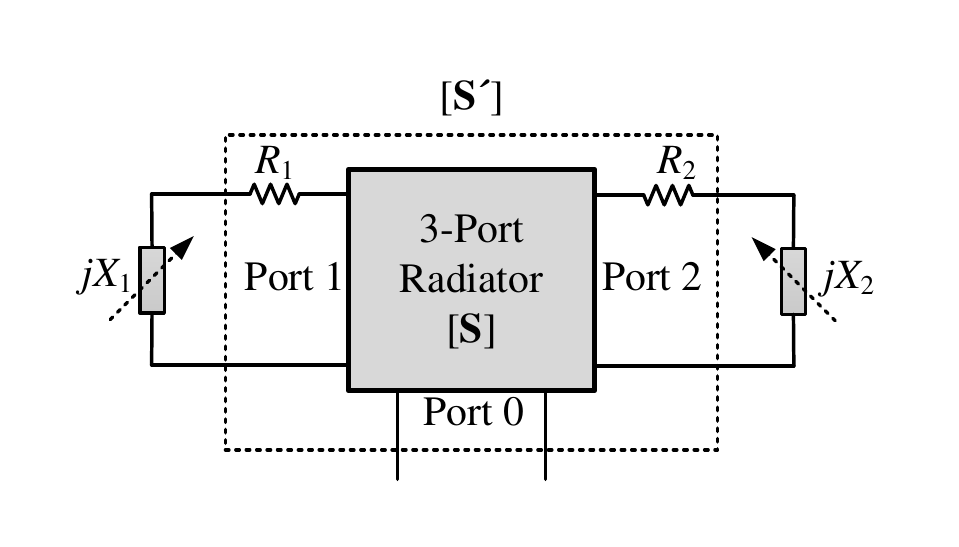}
\caption{The resistive loss ($R_1$ and $R_2$) of each variable load ($jX_1$ and $jX_2$) are included by introducing these in the scattering matrix of the three-port radiator.}
\label{fig:S_S}
\end{figure}

Since all the derivations in Section~\ref{secIIA} considered the use of purely reactive variable loads, the resistive loss of the varactor diodes is included in the scattering matrix of the three-port radiator, as depicted in Fig.~\ref{fig:S_S}, in order to maintain the accuracy of the calculations. In this case, the amended S-matrix can be expressed as 
%%%%%%%%%%%%%%%%%%%%%%%%%%%%%%
\begin{align} \label{eq:s2s}
\mathbf{S'} &= \left( {{\mathbf{Z'}} - {Z_0}{{\mathbf{I}}_3}} \right){\left( {{\mathbf{Z'}} + {Z_0}{{\mathbf{I}}_3}} \right)^{ - 1}} \\
&= \left[ {{\mathbf{Z}} + {\rm{diag}}(0,R_1,R_2) - {Z_0}{{\mathbf{I}}_3}} \right] \nonumber\\
&\quad\;\,{\left[ {{\mathbf{Z}} + {\rm{diag}}(0,R_1,R_2)+ {Z_0}{{\mathbf{I}}_3}} \right]^{ - 1}}\nonumber\\
& = \left[ {{Z_0}\left( {{{\mathbf{I}}_3} + {\mathbf{S}}} \right){{\left( {{{\mathbf{I}}_3} - {\mathbf{S}}} \right)}^{ - 1}} + {\rm{diag}}(0,R_1,R_2) - {Z_0}{{\mathbf{I}}_3}} \right]\nonumber\\
&\quad\;\,{\left[ {{Z_0}\left( {{{\mathbf{I}}_3} + {\mathbf{S}}} \right){{\left( {{{\mathbf{I}}_3} - {\mathbf{S}}} \right)}^{ - 1}} + {\rm{diag}}(0,R_1,R_2) + {Z_0}{{\mathbf{I}}_3}} \right]^{ - 1}} \nonumber 
\end{align}
%%%%%%%%%%%%%%%%%%%%%%%%%%%%%
where $\mathbf{Z}$ and $\mathbf{Z'}$ denote the impedance matrices of the three-port radiator before and after including the losses, and $R_1$ and $R_2$ are the average resistance values of the variable loads across the tuning range.

%%%%%%%%%%%%%%%%%%%%%%%%%%%%%%%%%%%%%%
%%%%%%%%%%%%%%%%%%%%%%%%%%%%%%%%%%%%%%%%%%

\begin{figure}[!t]
\centering
\subfigure[]{\includegraphics[width=3.0in]{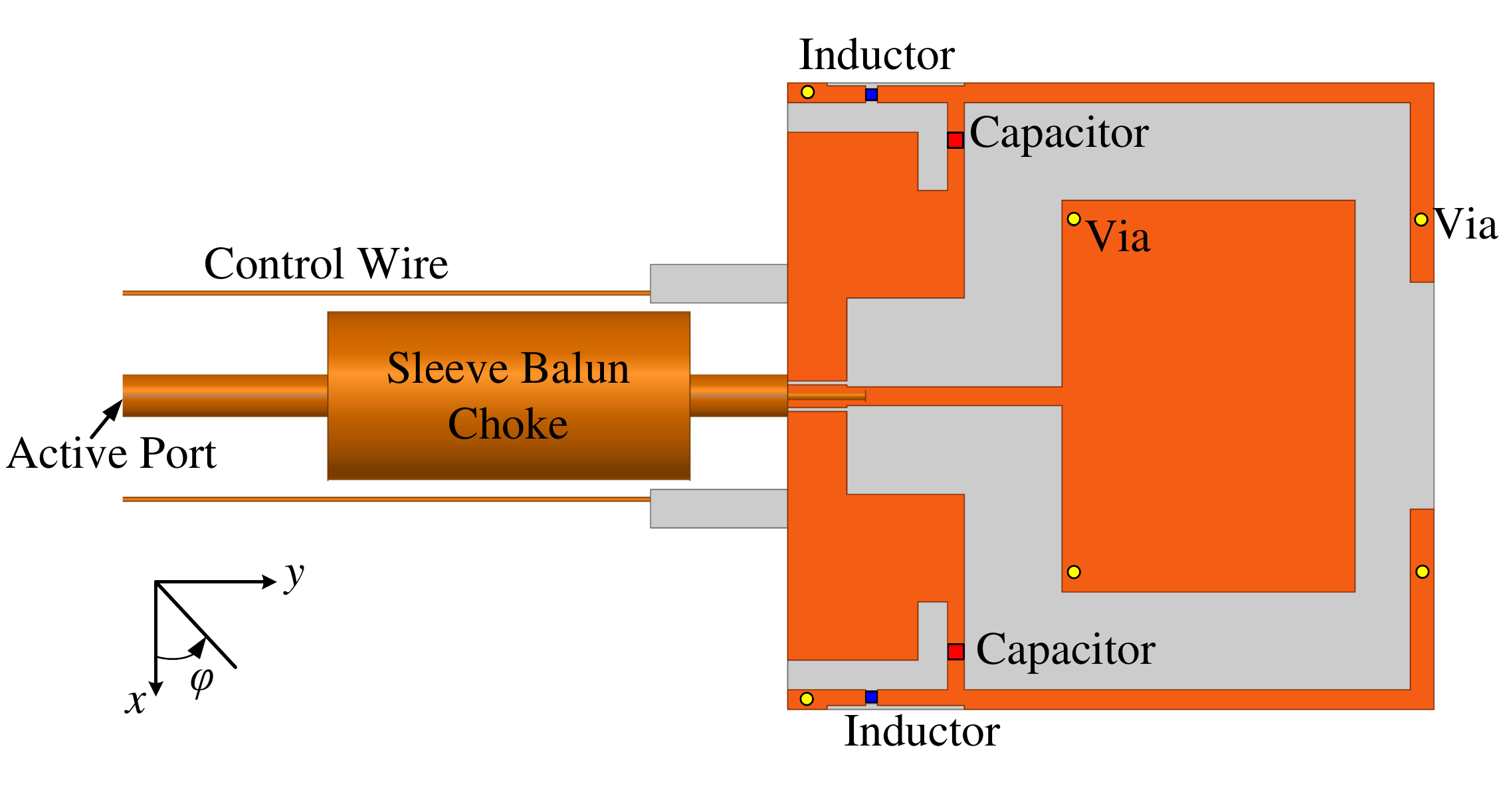} \label{fig:antenna_top}}
\subfigure[]{\includegraphics[width=3.0in]{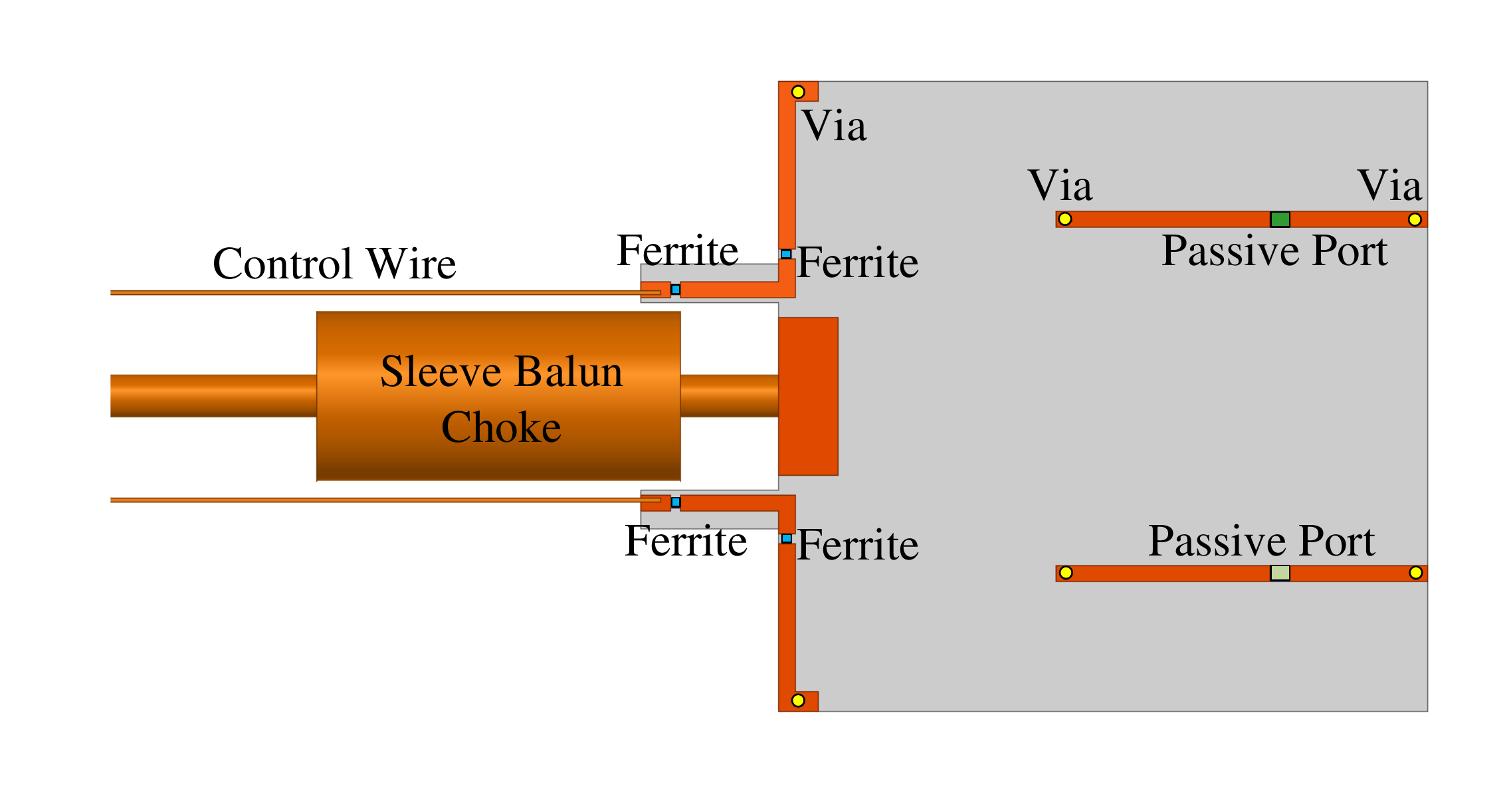} \label{fig:antenna_bot}}
\caption{Proposed three-port radiator with an axis of symmetry in \textit{yz}-plane. The substrate area is equivalent to $0.07\lambda_0^2$ at the design frequency. The quarter-wavelength balun acts as an open-end termination at the edge of the antenna. (a) Top view. (b) Bottom view.}
\label{fig:3port_radiator}
\end{figure}

As shown in the flowchart in Fig.~\ref{fig:flowchart}, the three-port radiator design procedure is started after determining the the variable loads. Fig.~\ref{fig:3port_radiator} shows the proposed radiator, which is realized on a $30$-mil-thick FR4 substrate of $32$~mm~$\times\;33$~mm, with a dielectric constant of $\epsilon_r = 4.4$. The structure is inspired from a folded slot antenna, reshaped by introducing two symmetric passive ports to bridge the slot. The central port is matched for a $50$-$\Omega$ coaxial connection to the single RF module. The other two ports are terminated with the varactor diodes. Two DC-decoupling $56$-pF capacitors (Murata GRM1885C2A560J) and two RF-blocking $56$-nH (Murata LQW15AN56NG) inductors are embedded in the design of the radiator to filter unwanted RF and control signals respectively. The bias voltages of the varactor diodes are applied between the control wires and the inner conductor of the coaxial cable.

An important challenge in the design of radiating structures with a small ground plane relates to correct excitation. In practice, a feeding cable is typically used to connect the antenna to other parts of the system or the measurement setup. However, if the ground plane of the antenna is small, current can flow back to the outer surface of the feeding cable, causing the cable itself to contribute to radiation and thereby change the input impedance. This introduces significant uncertainties in the result of full-wave simulations, where the antenna is fed using a lumped excitation or a finite-length feeding cable. To suppress unwanted cable currents, a standard sleeve balun choke was designed and placed in the path of the coaxial cable, very close to the radiator. 

%%%%%%%%%%%%%%%%%%%%%%%%%%%%
%%%%%%%%%%%%%%%%%%%%%%%%%%%%%%%%%%

Similarly, the wires used to convey the control waveforms of the varactor diodes are also in the near field of the radiator, and may re-radiate the induced currents. To prevent such undesired radiation, four chip ferrite beads (Murata BLM15GA750), with very high RF impedance, were placed in the path of the wires, as shown in Fig.~\ref{fig:3port_radiator}. To precisely model the embedded lumped components, equivalent circuit models were measured and corrected using the method presented in~\cite{my_apmag}. Table~\ref{tab:tab1} details the equivalent circuit models used in the simulations. 

%%%%%%%%%%%%%%%%%%%%%%%%%%%%%%%%%%%%
\begin{table}[t]
\caption{Corrected Equivalent Circuit Models}
\vspace{-2mm}
\label{tab:tab1}
\renewcommand{\arraystretch}{0.8}
\centering
% Some packages, such as MDW tools, offer better commands for making tables
% than the plain LaTeX2e tabular whI\!IEEEtranch is used here.
\footnotesize
\begin{tabular}{>{\centering}m{2.1cm}>{\centering}m{1.8cm}>{\centering}m{1.1cm}>{\centering}m{1.1cm}>{\centering}m{1.1cm}}
\toprule %\hline \hline \\[-1mm]
Component & Circuit Model  &  R ($\Omega$) & L (H) & C (F)\\
\midrule %\hline \\[-1mm]
capacitor $56$ pF & Series RLC & $1.0$ & -- & $100.0$ p\\[1mm]
inductor $56$ nH & Parallel RLC & $35.5$ k & $49.5$ n & $40.8$ f\\[1mm]
ferrite chip & Parallel RLC & $1.9$ k & $128.2$ n & $32.0$ f\\%[1mm]        
\bottomrule %\hline \hline   
\end{tabular}
\end{table}
%%%%%%%%%%%%%%%%%%%%%%%%%%%%%%%%%%%%%%%%%

Full-wave analyses of the three-port radiator were performed using ANSYS HFSS. As described in Fig.~\ref{fig:flowchart}, at each iteration step of the radiator design, the S-matrix and the embedded radiation patterns were exported to MATLAB and used to evaluate the specifications of the associated beam-space MIMO antenna, i.e., the input impedance, the radiation efficiency, the basis power imbalance ratio, and the required reactance values. A number of full-wave simulations were carried out to examine the effects of various key physical parameters of the radiator, and finally to arrive at a design fulfilling the requirements. 
% ACMA: can we say something more about how the design was iterated? was this some sort of GA/computer-based optimisation, or did you manually change things (if so, can we elaborate a little more on how you did this?)

\begin{figure}[!t]
\centering
\includegraphics[width=3.3in]{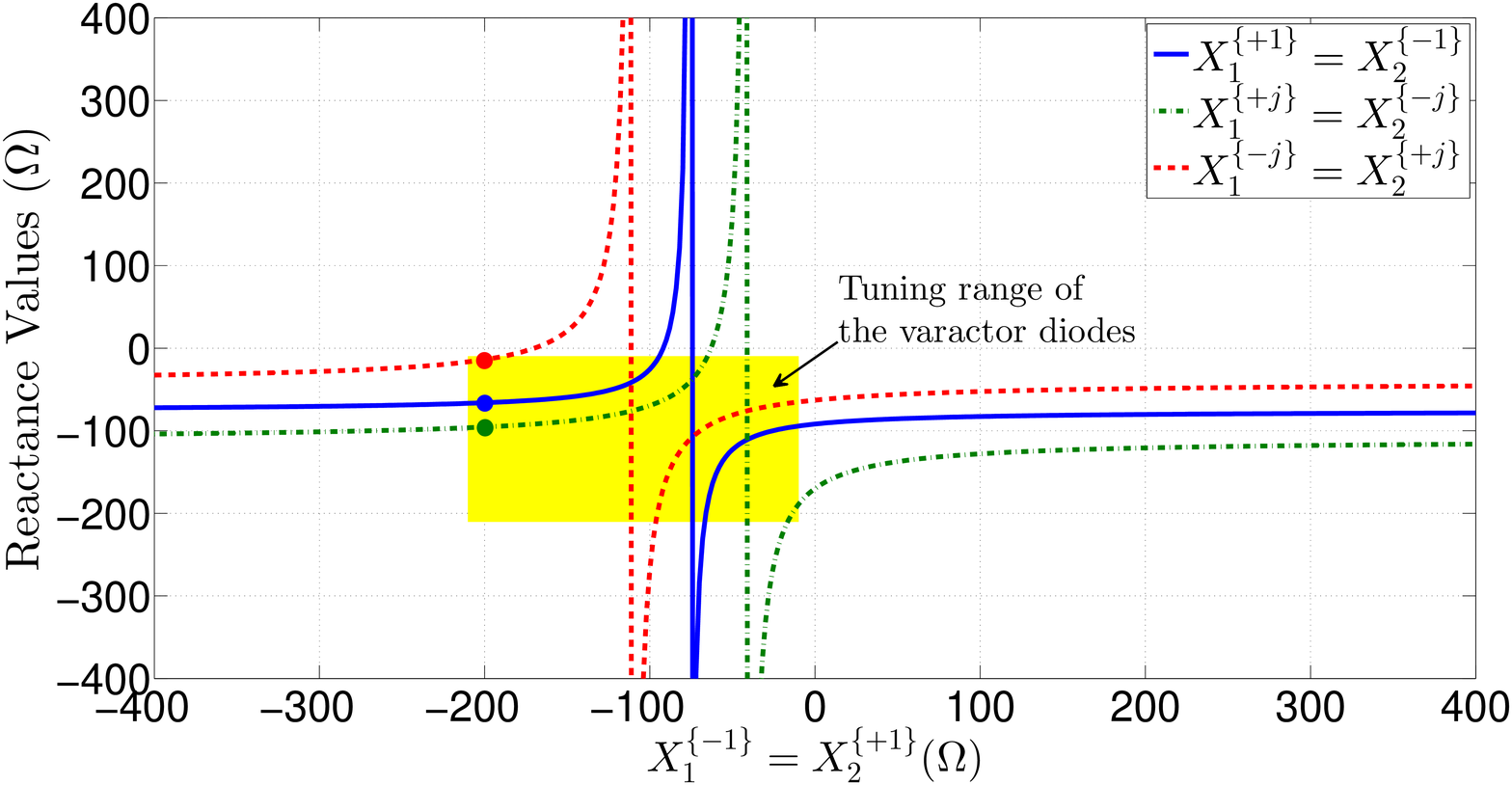}
\caption{Required reactance values as a function of the free parameter $X_1^{\{ -1\}}$ obtained from (\ref{eq:reactances}) and (\ref{eq:s2s}). The possible tuning range of the varactor diodes is shown by the shaded region.}
\label{fig:Xsolutions}
\end{figure}

The optimized three-port radiator, shown in Fig.~\ref{fig:3port_radiator}, yields an S-matrix of
%%%%%%%%%%%%%%%%%%%%%%%%%%%%%
\begin{equation*}\label{eq:Smatrix}
{\mathbf{S}} = \left[ {\begin{array}{*{20}{c}}
{ - 0.23 - j0.32}&{0.26 + j0.43}&{0.26 + j0.43}\\
{0.26 + j0.43}&{0.16 + j0.49}&{ - 0.19 - j0.11}\\
{0.26 + j0.43}&{ - 0.19 - j0.11}&{0.16 + j0.49}
\end{array}} \right]
\end{equation*}
%%%%%%%%%%%%%%%%%%%%%%%%%%%%%
at the design frequency. To complete the QPSK beam-space MIMO antenna, an arbitrary set of four reactance values are selected from the curves given in Fig.~\ref{fig:Xsolutions} to give a basis power imbalance ratio of $0.8$~dB and a return loss of $19.6$~dB. As shown in Fig.~\ref{fig:Xsolutions}, there exists a range of the solutions where all four reactance values lie within the specified tuning range of the varactor diodes. From this range, we selected the set associated with $X_1^{\{ -1\}} = -200\;\Omega$. Table~\ref{tab:tab2} gives the set of reactance values and their corresponding bias voltages.

%%%%%%%%%%%%%%%%%%%%%%%%%%%%%%%%%%%%%%
%%%%%%%%%%%%%%%%%%%%%%%%%%%%%%%%%%%%%%%%%%
%%%%%%%%%%%%%%%%%%%%%%%%%%%%%%%%%%%%
\begin{table}[!t]
\caption{Selected Reactance Set and Bias Voltages}
\vspace{-2mm}
\label{tab:tab2}
\renewcommand{\arraystretch}{0.8}
\centering
% Some packages, such as MDW tools, offer better commands for making tables
% than the plain LaTeX2e tabular whI\!IEEEtranch is used here.
\footnotesize
\begin{tabular}{>{\centering}b{0.8cm}>{\centering}b{0.6cm}>{\centering}b{1.5cm}>{\centering}b{1.5cm}>{\centering}b{1.5cm}>{\centering}b{1.5cm}}
\toprule %\hline \\[-1mm]
State & $\bar{x}$  &  $X_1^{\{ \bar{x}\}}$ ($\Omega$) & $X_2^{\{ \bar{x}\}}$ ($\Omega$) & $V_{\rm{bias},1}$ (V)& $V_{\rm{bias},2}$ (V)\\%[1mm]
\midrule %\\[-1mm]
$1$ & $-1$ & $-200.0$ & $-66.0$ & $-17.83$ & $-5.11$\\[1mm]
$2$ & $+1$ & $-66.0$ & $-200.0$ & $-5.15$ & $-17.78$\\[1mm]
$3$ & $+j$ & $-95.4$ & $-13.8$ & $-7.92$ & $-0.94$\\[1mm]
$4$ & $-j$ & $-13.8$ & $-95.4$ & $-0.75$ & $-7.40$\\%[1mm]
\bottomrule
\end{tabular}
\end{table}
%%%%%%%%%%%%%%%%%%%%%%%%%%%%%%%%%%%%%%%%%
%%%%%%%%%%%%%%%%%%%%%%%%%%%%%%%%%%%%%%
\begin{figure}[!t]
\centering
\includegraphics[width=2.0in]{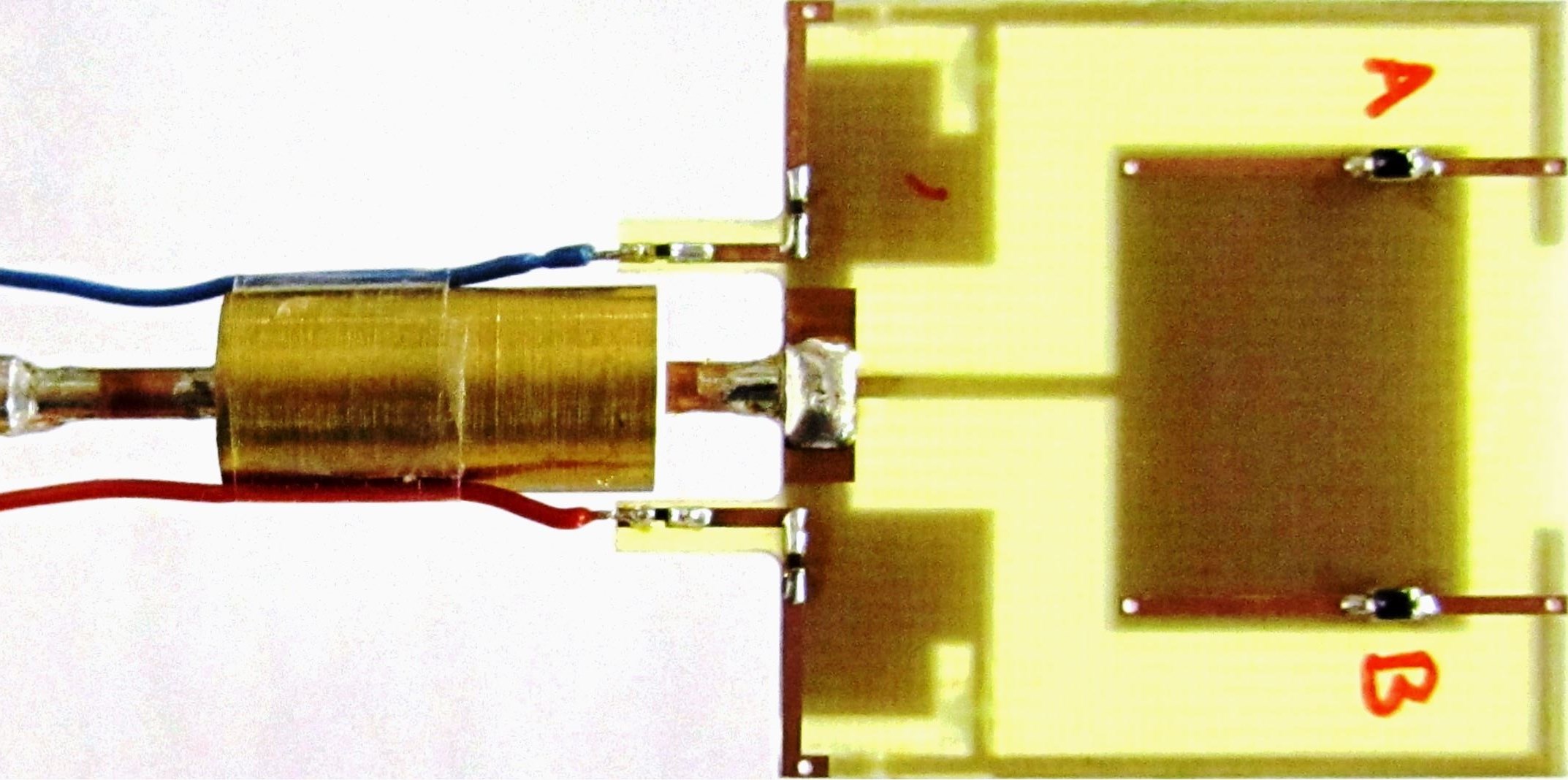}
\caption{Fabricated QPSK beam-space MIMO antenna.}
\label{fig:fab_ant}
\end{figure}
%%%%%%%%%%%%%%%%%%%%%%%%%%%%%%%%%%%%%%%%%%

\begin{figure}[!t]
\centering
\includegraphics[width=3.3in]{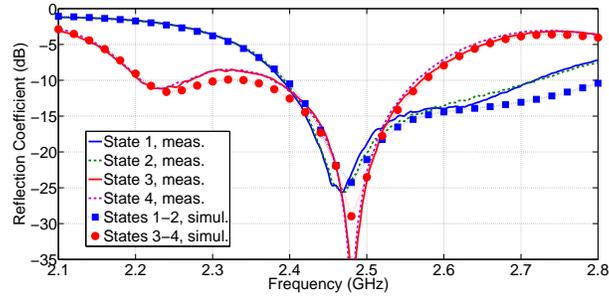}
\caption{Reflection coefficient of the QPSK beam-space MIMO antenna. We note that the operation bandwidth of the balun choke is narrow, and thus the simulation results may be inaccurate as the frequency deviates from the design frequency. }
\label{fig:RL}
\end{figure}

The optimized three-port radiator was fabricated and each of two varactor diodes under test were soldered to the corresponding passive port, as depicted in Fig.~\ref{fig:fab_ant}. The reflection coefficient of the antenna was measured for all four operational states, and compared to the results obtained from the simulations. For QPSK modulation  four antenna `states' are required and represent the instanteous radiation patterns of the beam-space MIMO antenna. Fig.~\ref{fig:RL} shows that the antenna yields a measured bandwidth of $7.0\%$ relative to the design centre frequency (for a reference of $-10$~dB) over all the states, while variations of the input impedance around the design frequency are negligible. Moreover, excellent agreement between the measured and simulated data validates the accuracy of the employed modeling approach.

%%%%%%%%%%%%%%%%%%%%%%%%%%%%%%%%%%%%%%
%%%%%%%%%%%%%%%%%%%%%%%%%%%%%%%%%%%%%%%%%%
%%%%%%%%%%%%%%%%%%%%%%%%%%%%
\begin{figure}[!t]
\centering
\subfigure[]{\includegraphics[width=4.4in]{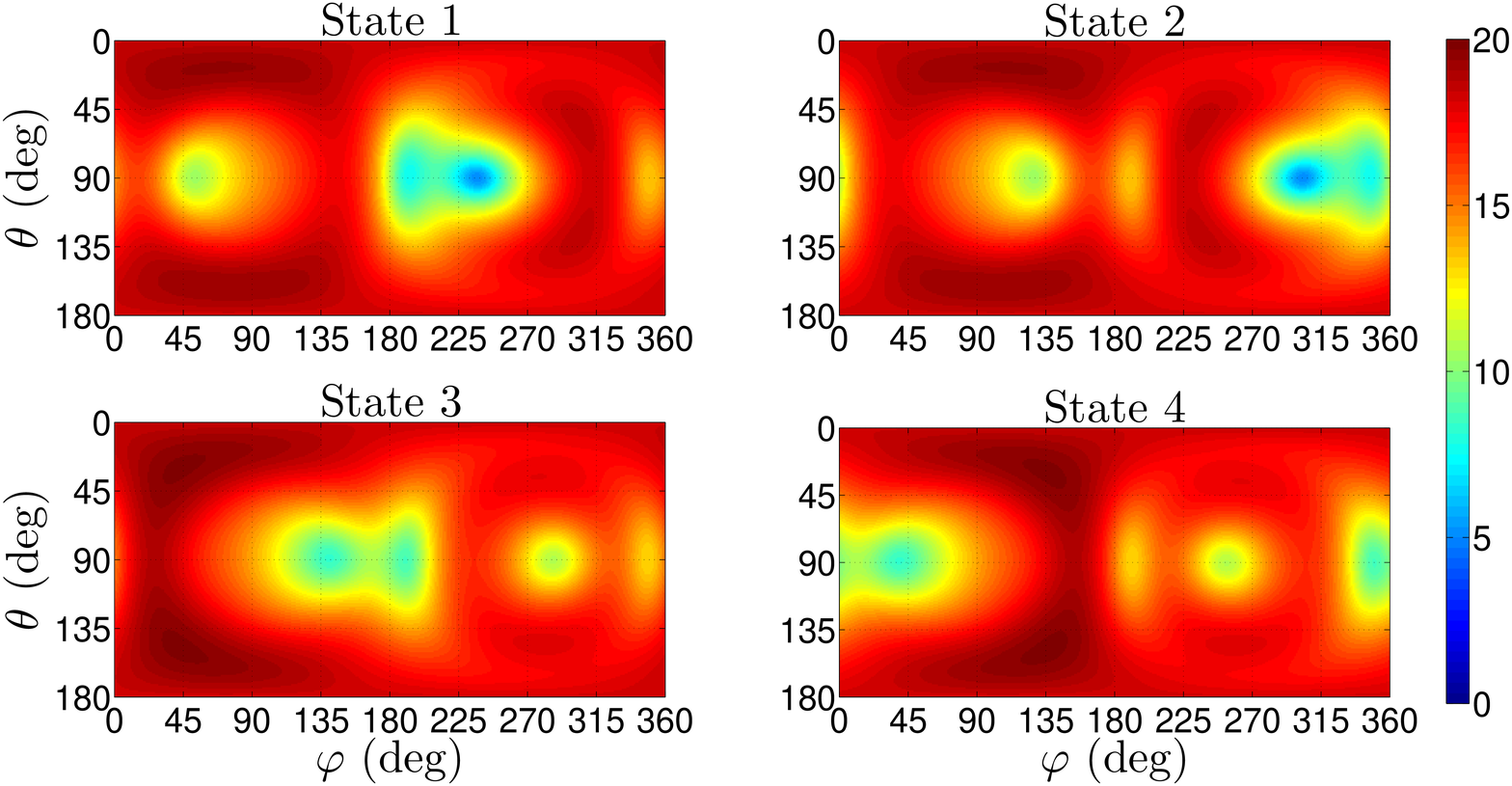} \label{fig:Eabs}}
\subfigure[]{\includegraphics[width=4.4in]{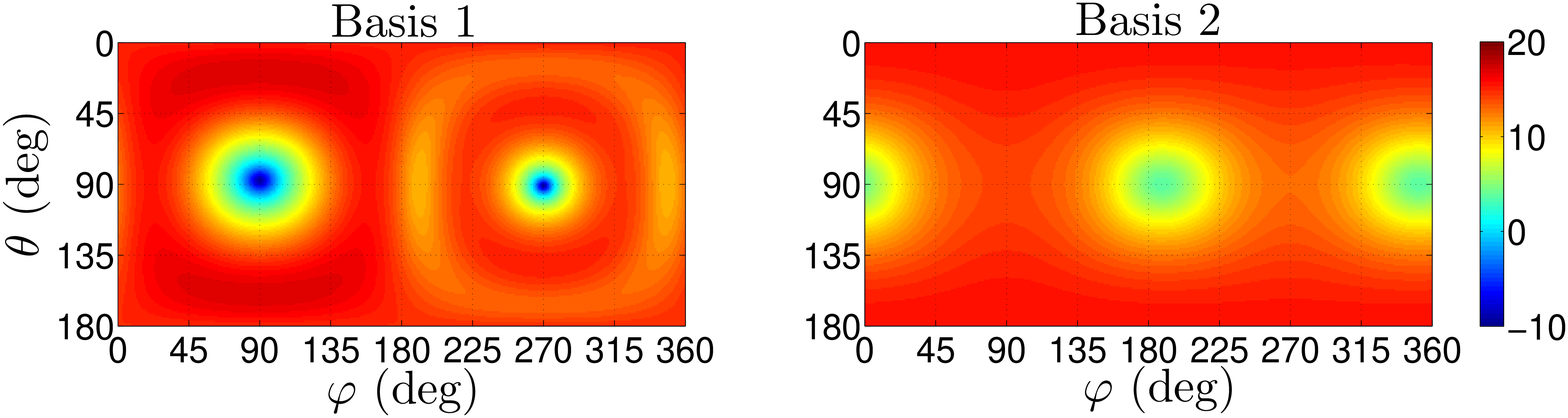} \label{fig:Babs}}
\caption{(a) Radiation patterns of the designed QPSK beam-space MIMO antenna at 2.45 GHz in dB. (b) Corresponding basis patterns.}
\label{fig:E_B_plot}
\end{figure}
%%%%%%%%%%%%%%%%%%%%%%%%%%%%%%%%%%

Fig.~\ref{fig:E_B_plot} shows the real instantaneous and virtual basis radiation patterns of the antenna obtained from full-wave simulations. As expected from the radiator symmetry and the switching of the variable loads, the instantaneous patterns in states 1 and 2, (like the instantaneous patterns in states 3 and 4), form mirrored pattern pairs with respect to the plane of symmetry in the $y$-plane (i.e., $\varphi = 90^{\circ}-270^{\circ}$). The virtual antenna associated with the first basis is observed to have a bowl-like radiation pattern, while the second basis pattern is bi-directional and broadside. Considering the thermal loss in the dielectric substrate, metallic parts, and resistance of the lumped components, the antenna yields a radiation efficiency of $75\%$ over all states.

\begin{figure}[!t]
\centering
\includegraphics[width=3.3in]{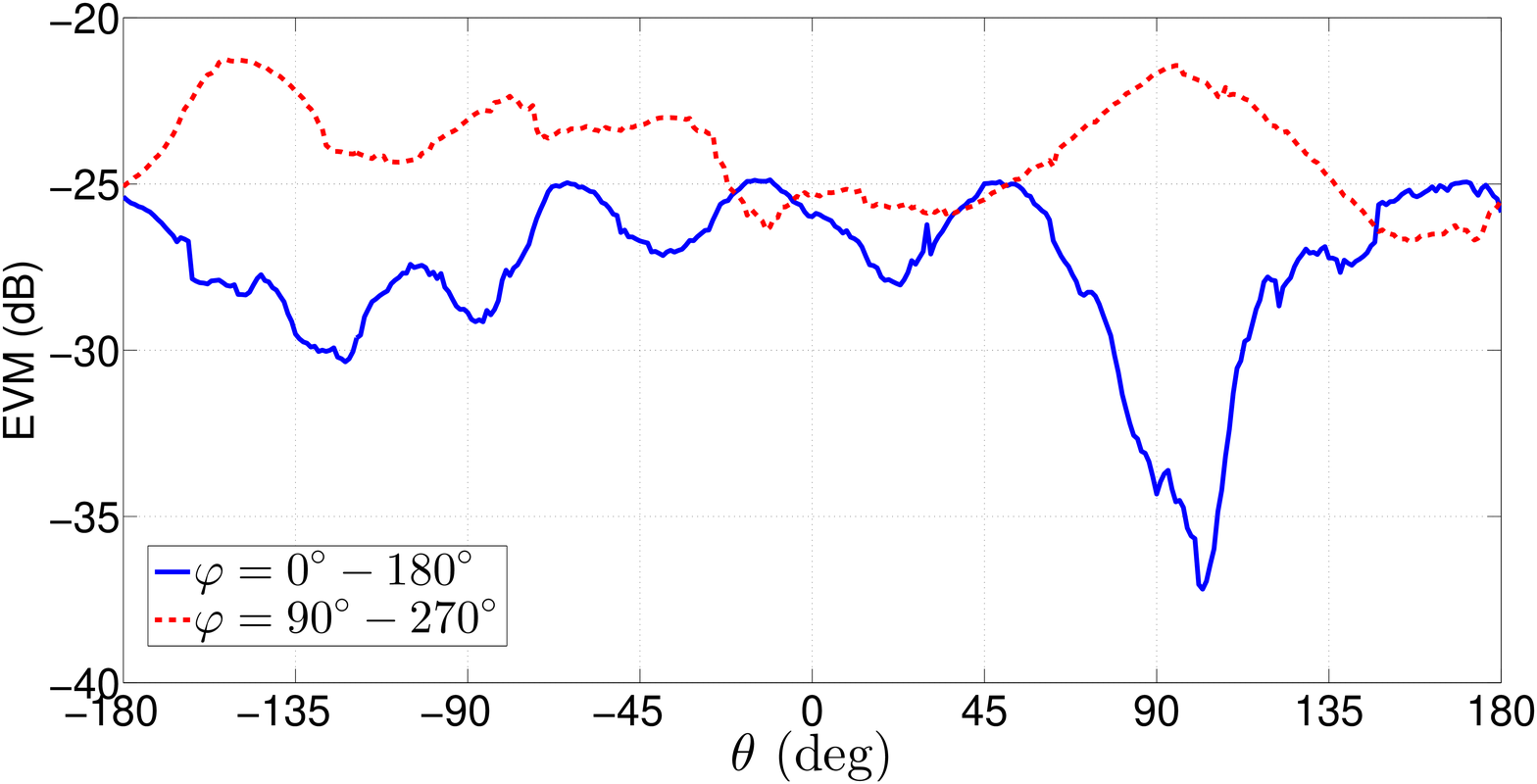}
\caption{Measured EVM of the QPSK beam-space MIMO antenna. The average of EVM is $-27.2$~dB and $-23.9$~dB in the planes of  $\varphi = 0^{\circ}-180^{\circ}$ and $\varphi = 90^{\circ}-270^{\circ}$, respectively. }
\label{fig:EVM}
\end{figure}

Since beam-space MIMO transmission is inherently conditional on proper and precise pattern reconfiguration, any imperfections in the antenna implementation (e.g., biasing imprecision, near-field coupling, and fabrication tolerances) may affect the modulation quality of the transmitted signals by distorting the location of the constellation points. This distortion can be quantified by the error vector magnitude (EVM)~\cite{EVM1}, which is reformulated for the fabricated QPSK beam-space MIMO antenna as
%%%%%%%%%%%%%%%%%%%%%%%%%%%%%%%%%%%%%%%%%%%%%%%%%%%%
\begin{equation}\label{eq:evm}
|{\rm{EVM}}(\theta ,\!\varphi )|^2 \!=\!  { {\frac{{\sum\limits_{k = 1}^4 \! {\| \boldsymbol{\mathcal{B}}_{1}(\theta ,\!\varphi ) \!+\! \bar{x}_{k} \boldsymbol{\mathcal{B}}_{2}(\theta ,\!\varphi ) \!-\! \boldsymbol{\mathcal{E}}_{\rm{unit}}^{\{\bar{x}_{k}\}}(\theta ,\!\varphi ) \|^2}}} {\sum\limits_{k = 1}^4 {\| \boldsymbol{\mathcal{B}}_{1}(\theta ,\varphi ) + \bar{x}_{k} \boldsymbol{\mathcal{B}}_{2}(\theta ,\varphi )  \|^2}}} } \,,
\end{equation}
%%%%%%%%%%%%%%%%%%%%%%%%%%%%%%%%%%%%%%%%%%%%%%%%%%%%%%
where $\{\bar{x}_{k}\}_{k=1}^4=\{-1,+1,+j,-j\}$ denote the four possible symbol combination ratios. To evaluate the EVM of the antenna, the complex radiation pattern was measured in an anechoic chamber, while exciting the antenna with a reference power. The measurements were carried out for all four antenna states and in two different planes. The results are shown in Fig.~\ref{fig:EVM}. The EVM of the fabricated antenna is less than $-20$~dB in both planes, revealing good modulation quality, despite residual imperfections in the antenna implementation and measurement procedures.
%%%%%%%%%%%%%%%%%%%%%%%%%%%%%%%%%%%%%%
%%%%%%%%%%%%%%%%%%%%%%%%%%%%%%%%%%%%%%%%%%
\section{Experimental Single-Radio MIMO Transmission}  \label{secIII}
%In Section~\ref{secII}, we designed a QPSK beam-space MIMO antenna, and verified its capability to create the required radiation patterns through far-field simulation and measurements. 
\textcolor{\markchanges}{In this section, we perform proof-of-concept experiments to demonstrate the concept of the beam-space multiplexing using the antenna designed in Section~\ref{secII}, and to compare its performance to conventional MIMO system with the same modulation scheme The general features of our over-the-air testbed are presented, followed by a description of the main design decisions concerning the transmitter and receiver subsystems. Finally, the results obtained with the testbed are presented.  
}

\textcolor{\markchanges}{
\subsection{Receiver Architecture for Beam-Space MIMO} \label{receiver_basics}
In beam-space MIMO, the radiation pattern of the transmit antenna is reconfigured at each symbol period, such that the symbols to be multiplexed are mapped onto the virtual basis patterns in the far-field. The signal transmitted by the beam-space MIMO antenna is then received using a multi-element antenna array at the receiving end. For clarity, we restrict our attention to the case where two PSK data streams are transmitted in the far-field from the PSK beam-space MIMO antenna proposed in Section~\ref{secII}, and received using a classical two-element array. The far-field embedded radiation patterns associated with the first and second receive antennas are  $\boldsymbol{\mathcal{F}}_{{\rm{R}},1}(\theta_{\rm{R}},\varphi_{\rm{R}})$ and $\boldsymbol{\mathcal{F}}_{{\rm{R}},2}(\theta_{\rm{R}},\varphi_{\rm{R}})$, respectively. According to the beam-space MIMO objective formulated in (\ref{eq:bsmimo_concept2}), the response of the $m$th receive antenna when illuminated by the radiation pattern $\boldsymbol{\mathcal{E}}_{{\rm{unit}}}^{\{\bar{x}\}}(\theta,\phi)$ can be expressed as 
%%%%%%%%%%%%%
\begin{align}\label{eq:hmn_inst}
h_m^{\{\bar{x}\}}&=\! \int{\!\!\!\!\!\int{\boldsymbol{\mathcal{F}}_{{\rm{R}},m}^{\rm{T}}({\Omega _{\rm{R}}}){\boldsymbol{\mathcal{O} }}({\Omega _{\rm{R}}},{\Omega})\boldsymbol{\mathcal{E}}_{{\rm{unit}}}^{\{\bar{x}\}}({\Omega}){\rm{d}}{\Omega}{\rm{d}}{\Omega _{\rm{R}}}}} \nonumber \\
&=\! \int{\!\!\!\!\!\int{\boldsymbol{\mathcal{F}}_{{\rm{R}},m}^{\rm{T}}({\Omega _{\rm{R}}}){\boldsymbol{\mathcal{O} }}({\Omega _{\rm{R}}},{\Omega })\left[ \boldsymbol{\mathcal{B}}_1({\Omega }) + \bar{x} \boldsymbol{\mathcal{B}}_2({\Omega })  \right]{\rm{d}}{\Omega }{\rm{d}}{\Omega _{\rm{R}}} }} \nonumber \\
&=\! \int{\!\!\!\!\!\int{\boldsymbol{\mathcal{F}}_{{\rm{R}},m}^{\rm{T}}({\Omega _{\rm{R}}}){\boldsymbol{\mathcal{O} }}({\Omega _{\rm{R}}},{\Omega })\boldsymbol{\mathcal{B}}_1({\Omega }) {\rm{d}}{\Omega }{\rm{d}}{\Omega _{\rm{R}}} }} \nonumber \\
& \;\; \;\; + \bar{x} \int{\!\!\!\!\!\int{\boldsymbol{\mathcal{F}}_{{\rm{R}},m}^{\rm{T}}({\Omega _{\rm{R}}}){\boldsymbol{\mathcal{O} }}({\Omega _{\rm{R}}},{\Omega }) \boldsymbol{\mathcal{B}}_2({\Omega })  {\rm{d}}{\Omega }{\rm{d}}{\Omega _{\rm{R}}} }} \nonumber \\
&= \hbar_{m1} + \bar{x} \hbar_{m2} \,,
\end{align}
%%%%%%%%%%%%%%%
where $\hbar_{m1}$ and $\hbar_{m2}$ are the responses of the $m$th receive antenna to the virtual basis patterns, $\boldsymbol{\mathcal{B}}_1(\Omega)$ and $\boldsymbol{\mathcal{B}}_2(\Omega)$, respectively; the solid angles at the transmitter and receiver are given by $\Omega$ and $\Omega_{\rm{R}}$, respectively; and the physical scattering (spatial spreading) taking into account path loss, polarization transformation, scattering, diffraction, and other channel effects are included in ${\boldsymbol{\mathcal{O} }}({\Omega _{\rm{R}}},{\Omega})$.}

\textcolor{\markchanges}{
Recalling that the first stream, $x_1$, is up-converted and used to excite the beam-space MIMO antenna, the baseband representation of the received signal vector at the two receiving antennas is  
%%%%%%%%%%%%%
\begin{align}\label{eq:yy1}
\mathbf{y}^{\{\bar{x}\}} &= x_1 \left[ {\begin{array}{*{20}{c}}
{h_1^{\{\bar{x}\}}}\\
{h_2^{\{\bar{x}\}}}
\end{array}} \right] + \mathbf{n} =  x_1 \left[ {\begin{array}{*{20}{c}}
{\hbar_{11} + \bar{x} \hbar_{12}}\\
{\hbar_{21} + \bar{x} \hbar_{22}}
\end{array}} \right] + \mathbf{n} \nonumber \\ 
&= \left[ {\begin{array}{*{20}{c}}
\hbar_{11} & \hbar_{12}\\
\hbar_{21} & \hbar_{22}
\end{array}} \right] \left[ {\begin{array}{*{20}{c}}
{x_1}\\
{x_2}
\end{array}} \right] + \mathbf{n} \,,
\end{align}
%%%%%%%%%%%%%%%
where $\mathbf{n}$ represents the receive noise vector, and $\{x_1,x_2\}$ are the assumed transmit symbols. Under ideal conditions, where the modulation on the orthogonal basis patterns is perfect, (\ref{eq:yy1}) is valid for all the possible values of $\bar{x}$. Therefore, the beam-space MIMO system model follows the conventional MIMO system model, i.e., 
%%%%%%%%%%%%%
\begin{equation}\label{eq:yy2}
\mathbf{y}=\mathbf{Hx}+\mathbf{n}\,,
\end{equation}
%%%%%%%%%%%%%%%
where 
%%%%%%%%%%%%%
\begin{equation}\label{eq:H1}
\mathbf{H}={\left[ {\begin{array}{*{20}{c}}
\hbar_{11} & \hbar_{12}\\
\hbar_{21} & \hbar_{22}
\end{array}} \right]}
\end{equation}
%%%%%%%%%%%%%%%
is the channel matrix, and the entries of $x$ are independently chosen from the scalar PSK constellation alphabet. Accordingly, a linear zero-forcing receiver, followed by component-wise quantization to the nearest constellation point, can recover the two transmitted streams.}

\subsection{Testbed Setup}  \label{sec_testbedSetup}
%To build a real-time testbed, we combined the NI software defined radio platforms and the LabVIEW communications system design software, providing a unified hardware and software design flow. 

%%%%%%%%%%%%%%%%%%%%%%%%%%%
\begin{figure}[!t] 
\centering 
\includegraphics[width=3.7in]{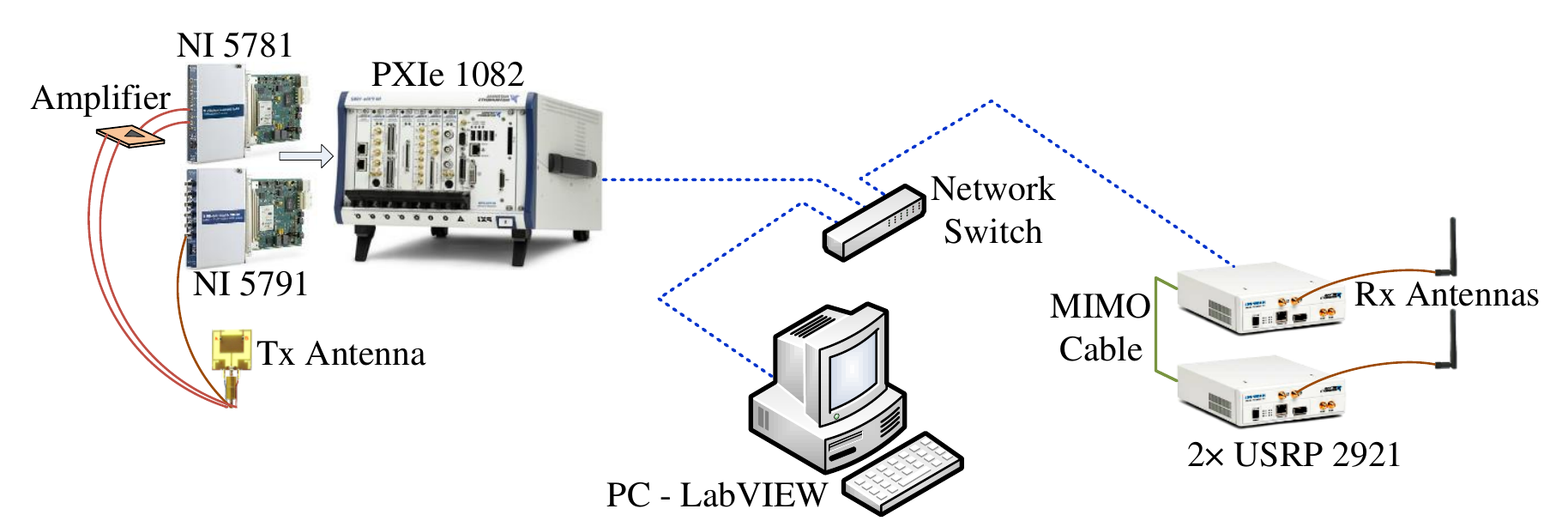} 
\caption{The hardware connectivity in the beam-space MIMO testbed.}
\label{fig:testbed1} 
\end{figure}
\begin{figure}[!t] 
\centering 
\includegraphics[width=3.4in]{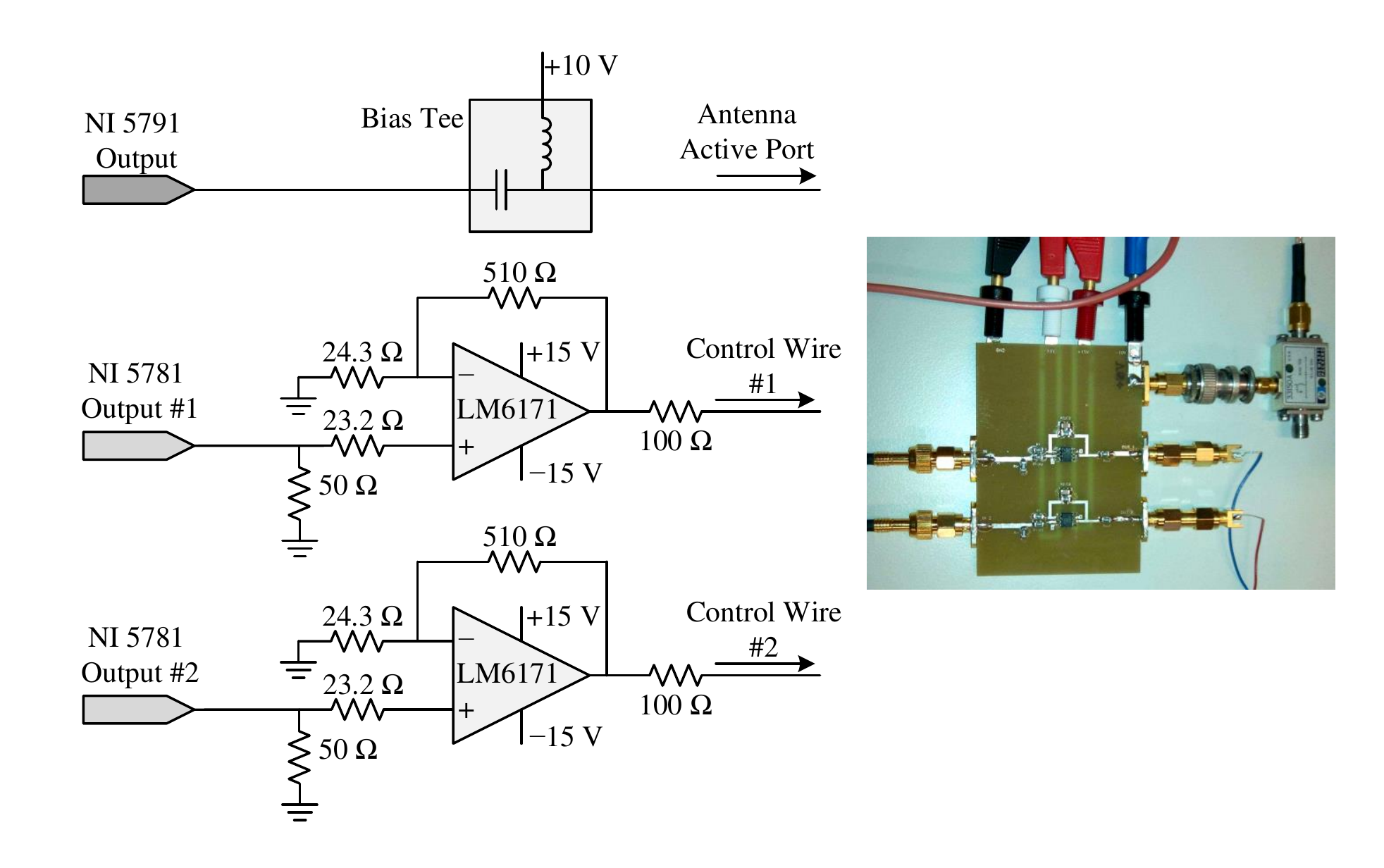}  
\caption{Amplifier circuit used for amplifying the baseband control signals.}
\label{fig:amp} 
\end{figure}

%%%%%%%%%%%%%%%%%%%%%%%%%%%%%
\textcolor{\markchanges}{
Fig.~\ref{fig:testbed1} shows a schematic of the testbed setup, where the transmitting and receiving subsystems were connected to the host computer via a network switch. At the transmitting end, we used the NI FlexRIO platform, which includes reconfigurable field-programmable gate arrays (FPGAs) for real-time processing and deterministic control. Specifically, one RF transceiver adapter module (NI~5791) and one baseband transceiver adapter module (NI~5781) were used to produce the required single RF signal (fed into the active port of the beam-space MIMO antenna) and the control signals (for controlling the states of the variable loads), respectively. Both the RF and baseband modules were connected to an NI PXIe-1082 real-time host, and controlled by NI FlexRIO FPGA modules.} The voltage dynamic range at the outputs of the baseband NI~5781 module is limited to $1$~V peak-to-peak (i.e., between $\pm 0.5$~V), however, as shown in Table~\ref{tab:tab2} a bias voltage range from $-18$~V to $0$~V is required for the varactor diodes embedded in the antenna. Therefore, an amplifier circuit was used to increase the voltage range of the load-control signals. Fig.~\ref{fig:amp} shows the amplifier design, implemented using two LM6171 high-speed, low-power, low-distortion feedback amplifiers from Texas Instruments. To achieve an asymmetric tuning range (namely, from $-20$~V to $0$~V rather than from $-10$~V to $+10$~V), a constant voltage offset of $+10$~V was applied through the inner conductor of the coaxial cable using a bias tee to the common side of the varactor diodes. \textcolor{\markchanges}{Table~\ref{tab:Vamp} gives the required voltage values at the amplifier outputs for the different antenna states.}

%%%%%%%%%%
%%%%%%%%%%%
%%%%%%%%%%%%%%%%%%%%%%%%%%%%%%%%%%%%
\begin{table}[!t]
\caption{Actual Bias Voltages Versus Voltages at Amplifier Outputs}
\vspace{-2mm}
\label{tab:Vamp}
\renewcommand{\arraystretch}{0.8}
\centering
% Some packages, such as MDW tools, offer better commands for making tables
% than the plain LaTeX2e tabular whI\!IEEEtranch is used here.
\footnotesize
\begin{tabular}{>{\centering}b{0.65cm}>{\centering}b{1.75cm}>{\centering}b{1.75cm}>{\centering}b{1.75cm}>{\centering}b{1.75cm}}
\toprule %\hline \hline \\[-1mm]
State &  $V_{\rm{bias},1}$ (V)& $V_{\rm{bias},2}$ (V) & $V_{\rm{amp},1}$ (V)& $V_{\rm{amp},2}$ (V)\\%[1mm]
\midrule %\hline \\[-1mm]
$1$  &  $-17.83$ & $-5.11$ & $-7.83$ & $4.89$\\[1mm]
$2$  & $-5.15$ & $-17.78$ & $4.85$ & $-7.78$\\[1mm]
$3$  & $-7.92$ & $-0.94$ & $2.08$ & $9.06$\\[1mm]
$4$  & $-0.75$ & $-7.40$ & $9.25$ & $2.60$\\%[1mm]
\bottomrule %\hline\hline  
\end{tabular}
\end{table}
%%%%%%%%%%%%%%%%%%%%%%%%%%%%%%%%%%%%%%%%%
\textcolor{\markchanges}{
At the receiving end, we employed two standard $2.4$~GHz `rubber duck' antennas, each connected to an NI Universal Software Radio Peripheral (USRP-2921) device. A MIMO cable was used to synchronize the USRP pair in time and frequency. Unlike the transmitter, the receiver did not run on a real-time host. Instead, as shown in Fig.~\ref{fig:testbed1}, the USRP devices were connected to the host computer where the baseband signal processing operations were performed.  
}

\textcolor{\markchanges}{
\subsection{System Design and Functionality} \label{sec_ota_bsmimo}
%%%%%%%%%%
\begin{figure*}
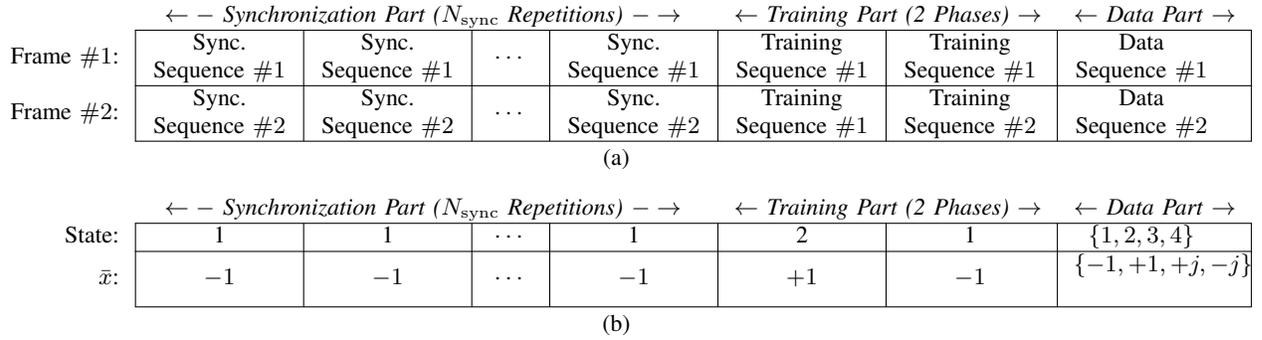

\addtolength{\subfigcapskip}{1mm}
\centering
{
\footnotesize
\subfigure[]{
\begin{tabular}{>{\raggedleft}m{1.5cm}|>{\centering}m{1.8cm}|>{\centering}m{1.8cm}|>{\centering}m{0.6cm}|>{\centering}m{1.8cm}|>{\centering}m{1.8cm}|>{\centering}m{1.8cm}|>{\centering}m{1.8cm}|}
\multicolumn{1}{c}{}    & \multicolumn{4}{c}{$\leftarrow -$ \emph{Synchronization Part (${N}_{\rm{sync}}$ Repetitions)} $-\rightarrow$ } & \multicolumn{2}{c}{$\leftarrow$  \emph{Training Part (2 Phases)} $\rightarrow$} & \multicolumn{1}{c}{$\leftarrow$  \emph{Data Part}  $\rightarrow$}\\
\cline{2-8}
Frame $\#1$: & Sync. Sequence $\#1$ & Sync. Sequence $\#1$ & $\cdots$ & Sync. Sequence $\#1$ & Training Sequence $\#1$ & Training Sequence $\#1$ & Data Sequence $\#1$  \\ 
\cline{2-8}
Frame $\#2$: & Sync. Sequence $\#2$ & Sync. Sequence $\#2$ & $\cdots$ & Sync. Sequence $\#2$ & Training Sequence $\#1$ & Training Sequence $\#2$ & Data Sequence $\#2$   \\ 
\cline{2-8}
\end{tabular} 
}
\\
\subfigure[] {
\footnotesize
\begin{tabular}{>{\raggedleft}m{1.5cm}|>{\centering}m{1.8cm}|>{\centering}m{1.8cm}|>{\centering}m{0.6cm}|>{\centering}m{1.8cm}|>{\centering}m{1.8cm}|>{\centering}m{1.8cm}|>{\centering}m{1.8cm}|}
\multicolumn{1}{c}{}    & \multicolumn{4}{c}{$\leftarrow -$ \emph{Synchronization Part (${N}_{\rm{sync}}$ Repetitions)} $-\rightarrow$ } & \multicolumn{2}{c}{$\leftarrow$  \emph{Training Part (2 Phases)} $\rightarrow$} & \multicolumn{1}{c}{$\leftarrow$  \emph{Data Part}  $\rightarrow$}\\
\cline{2-8}
 State: & 1  & 1  & $\cdots$  & 1  & 2 & 1 & $\{1,2,3,4\}$    \\ 
\cline{2-8} 
$\bar{x}$: & $-1$ & $-1$ & $\cdots$ & $-1$ & $+1$ & $-1$ & $\{-1,+1,+j,-j\}$ \\ 
\cline{2-8} 
\end{tabular} 
}
}
\caption{(a) Frame structure used for beam-space MIMO experiments. (b) Corresponding antenna state and the symbol combination ratio, $\bar{x}$, during the frame transmission.}
\label{fig:bsmimo_frame1}
\end{figure*}
%%%%%%%%%%%%
%To perform an over-the-air experiment, the main software calls the transmission and reception programs. 
%The transmission program, which is compiled and deployed on the host machine, supports a frame-based transmission. 
%The procedure starts by generating two frames of symbols for single-radio MIMO transmission. 
The MIMO system implemented on the testbed is frame-based. The symbols of the first frame (representing the RF path and directly applied to the antenna) are oversampled and pulse-shaped with a root raised cosine filter and up-converted to the carrier frequency. Simultaneously, in the baseband path, the control signals are generated according to the ratio of the symbols in each frame. Fig.~\ref{fig:bsmimo_frame1} illustrates the structure of the transmission frames. The first part of each frame includes ${N}_{\rm{sync}}$ repetitions of a pseudo-random synchronization sequence, which is used at the receiver for timing acquisition and the estimation of the frequency offset between the transmitter and the receiver~\cite{freq_off}. The synchronization sequence of the second frame is set (with respect to the first frame), such that the symbol combination ratio, $\bar{x}$, is constant and equal to $-1$, thereby forcing the antenna to radiate in state $1$. Following the synchronization sequences, each frame contains two training sequences, which are used at the receiver to estimate the channel matrix. The first sequence in each frame is identical to guarantee a symbol combination ratio of $+1$ during the first phase of the training. However, similar to the synchronization stage, the second training sequence is set such that the symbol combination ratio is equal to $-1$. The remainder of each frame is dedicated to (randomly generated) data. 
}
%ACMA: can we expand a little on why it is important to use different antenna states to do the channel estimation

% based on Table~\ref{tab:Vamp}.  
\textcolor{\markchanges}{
Since the variable loads of the antenna need to be modulated at the symbol rate of the RF channel, the control signals are also oversampled with the same factor as the RF signal. The RF and baseband samples are then uploaded to their corresponding FPGAs. A network communication link is established between the transmitter and the receiver to send the system parameters, the synchronization and training sequences, and the matched filter coefficients. 
After triggering the two transceiver adapter modules at the same time, the RF and baseband control samples are supplied to the NI~5791 and NI~5781 modules, respectively. A handshaking protocol between the transmitter and the receiver is used to ensure that the frame transmission occurs within the observation window of the receiver. 
}

\textcolor{\markchanges}{
The host computer collects the received data samples from the USRP devices and performs conventional signal processing operations (i.e., frequency offset estimation and correction, timing synchronization, channel estimation, and zero-forcing equalization) in order to decouple the two QPSK symbol streams. According to (\ref{eq:yy2}) and (\ref{eq:H1}), under ideal conditions the first and second training phases stimulate  
%%%%%%%%%%%%%%%%%%%%%%%
\begin{align} \label{eq:y_training}
\mathbf{y}_{{\rm{t}},1} &= \left[\!\!{\begin{array}{*{20}{c}}
\hbar_{11}+\hbar_{12}\\
\hbar_{21}+\hbar_{22}
\end{array}} \!\!\right] \!\mathbf{t} + \mathbf{n}_1 = \left[\!\! {\begin{array}{*{20}{c}}
\hbar_{11}&\hbar_{12}\\
\hbar_{21}&\hbar_{22}
\end{array}} \!\!\right] \!\! \left[\!\! {\begin{array}{*{20}{c}}
\mathbf{t}\\
\mathbf{t}
\end{array}} \!\!\right] + \mathbf{n}_1 \\
\mathbf{y}_{{\rm{t}},2} &= \left[\!\!{\begin{array}{*{20}{c}}
\hbar_{11}-\hbar_{12}\\
\hbar_{21}-\hbar_{22}
\end{array}} \!\!\right] \!\mathbf{t} + \mathbf{n}_2 = \left[\!\! {\begin{array}{*{20}{c}}
\hbar_{11}&\hbar_{12}\\
\hbar_{21}&\hbar_{22}
\end{array}} \!\!\right] \!\! \left[\!\! {\begin{array}{*{20}{c}}
\mathbf{t}\\
-\mathbf{t}
\end{array}} \!\!\right] + \mathbf{n}_2 %\nonumber
\end{align}
%%%%%%%%%%%%%%%%%%%%%%%%%%%%%%%%%%%%%%%%%%%%%%%
at the two receive antennas, respectively, where $\mathbf{t}$ denotes the training sequence. Therefore, the channel matrix can be estimated using a least squares approach as
%%%%%%%%
\begin{equation} \label{eq:Hhat}
\mathbf{\widehat{H}}= \left[\!\! {\begin{array}{*{20}{c}}
\hat{\hbar}_{11}&\hat{\hbar}_{12}\\
\hat{\hbar}_{21}&\hat{\hbar}_{22}
\end{array}} \!\!\right] = \left[\!\! {\begin{array}{*{20}{c}}
\mathbf{y}_{t,1}&\mathbf{y}_{t,2}
\end{array}} \!\!\right]\mathbf{T}^{\rm{H}}\left(\mathbf{T}\mathbf{T}^{\rm{H}}\right)^{-1} \,,
\end{equation}
%%%%%%%%%%%%%
where   
%%%%%%%%%%%%%%%%%%%%%%%%%%%%%%%%%%%%%%%%%%%%%%%%%%%%%%%%%%%%%%%%%%%%%%%%%%%%%%%%%%%%%%%%
\begin{equation*}
\mathbf{T}=\left[\!\! {\begin{array}{*{20}{c}}
\mathbf{t}&\mathbf{t}\\
\mathbf{t}&-\mathbf{t}
\end{array}} \!\!\right]\,\cdot 
\end{equation*}
%%%%%%%%%%%%%%%%%%%%%%%%%%%%%%%%%%%%%%%%%%%%%
Once the channel matrix has been estimated, the zero-forcing decoder simply multiplies the decimated received signal vector with the inverse of the channel matrix.
}

\begin{figure}[!t] 
\centering 
\includegraphics[width=3.4in]{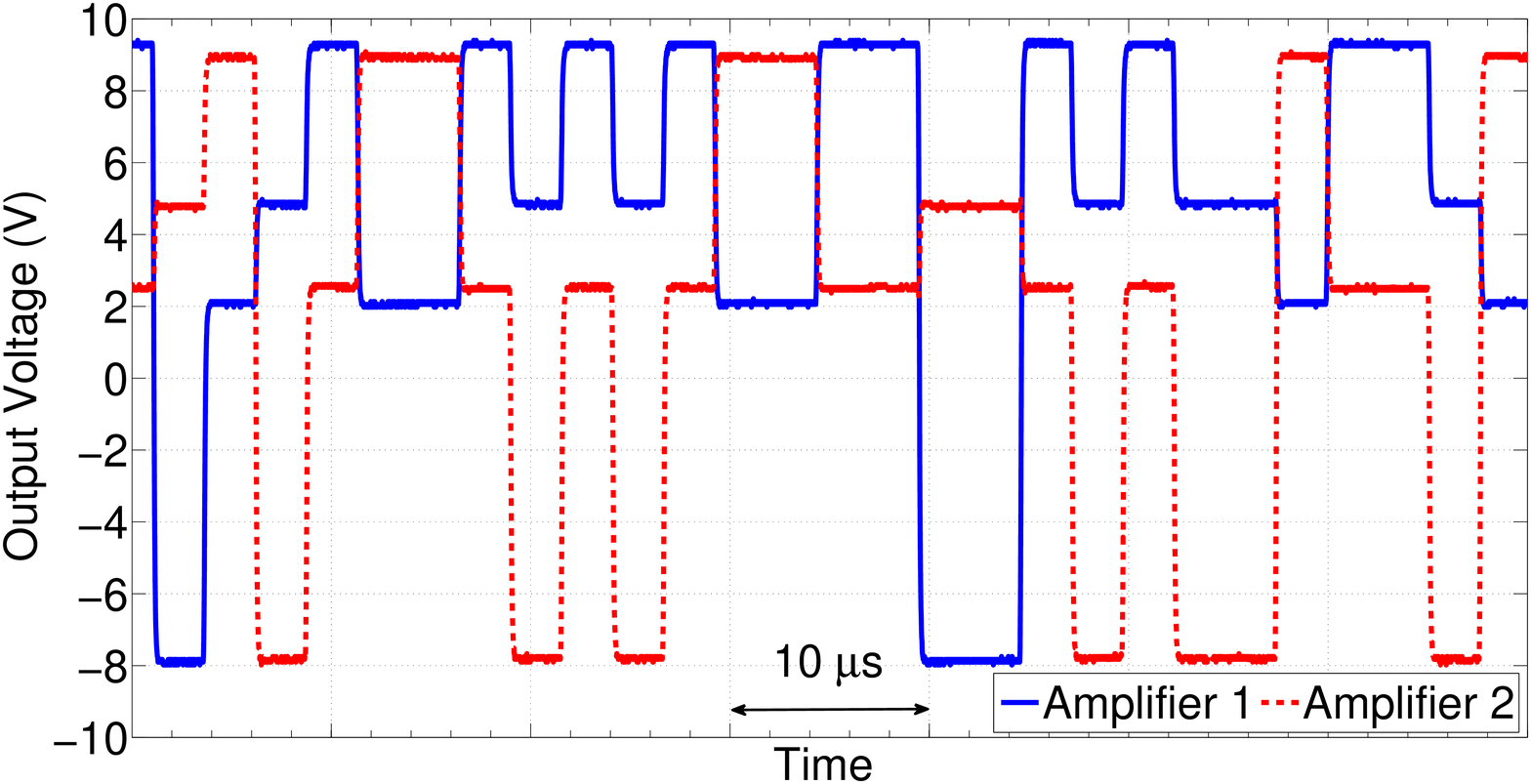} 
\caption{Baseband control signals (during the data segment) at the outputs of the amplifier.}
\label{fig:bbOUT} 
\end{figure}

\begin{figure}[!t] 
\centering 
\includegraphics[width=3.4in]{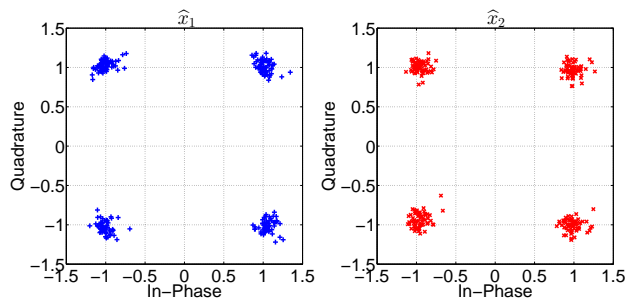} 
\caption{Scatter plot of received signal constellation after zero-forcing equalization.}
\label{fig:scatterplot1} 
\end{figure}

\textcolor{\markchanges}{
\subsection{Proof-of-Concept Results}
The over-the-air experiment was carried out in an indoor non-line-of-sight (NLOS) laboratory environment at EPFL. The key specifications of the testbed are summarized in Table~\ref{tab:parameters1}. Fig.~\ref{fig:bbOUT} depicts the control waveforms at the outputs of the amplifiers, where the voltage levels are in good agreement with the desired values in Table~\ref{tab:Vamp}.}  
Fig.~\ref{fig:scatterplot1} shows the recovered symbol constellations after zero-forcing equalization. It is clearly observed that the two QPSK streams were properly recovered. This experiment confirms the excellent behavior of the designed antenna, and constitutes the first successful single-radio MIMO transmission of QPSK signals. \textcolor{\markchanges}{These proof-of-concept results are extended in the following Section, to better quantify and compare the performance of the beam-space MIMO system with a conventional $2\times2$ MIMO system in real indoor environments.
}

%%%%%%%%%%%%%%%%%%%%%%%%%%%%%%%%%%%%
\begin{table}[!t]
\caption{System Parameters of the Validation Experiment}
\vspace{-2mm}
\label{tab:parameters1}
\renewcommand{\arraystretch}{0.8}
\centering
% Some packages, such as MDW tools, offer better commands for making tables
% than the plain LaTeX2e tabular whI\!IEEEtranch is used here.
\footnotesize
\begin{tabular}{>{\centering}b{2.5cm}>{\centering}b{2.5cm}>{\centering}b{2.5cm}>{\centering}b{2.5cm}}
\toprule %\hline \hline \\[-1mm]
$N_{\rm{sync}}$ &  Sync. Seq.  & Training Seq.  & Data Seq. \\[1mm]
$4$  &  $32$ symbols & $64$ symbols & $256$ symbols \\[1mm]
\midrule %\hline \\  
Frequency &  Tx Sampling Rate & Symbol Rate & Rx Sampling Rate \\[1mm]
$2.45$ GHz  &  $100$ MHz & $390.625$ kHz & $1562.5$ kHz \\%[1mm]
\bottomrule %\hline\hline 
\end{tabular}
\end{table}
%%%%%%%%%%%%%%%%%%%%%%%%%%%%%%%%%%%%%%%%%

%%%%%%%%%%%%%%%
%%%%%
%%%
%%%%%

%%%%%%%%%%%%%%%%%%%%%%%%%%%%%%%%%%%%%%%%%%%%%%%%%%%%%%%%%%%%%%%%%%%%%%%%%%%%%%%%%%%%%%%%%%%%%%%%
%%%%%%%%%%%%%%%%%%%%%%%%%%%%%%%%%%%%%%%%%%%%%%%%%%%%%%%%%%%%%%%%%%%%%%%%%%%%%%%%%%%%%%%%%%%%%%%%
%%%%%%%%%%%%%%%%%%%%%%%%%%%%%%%%%%%%%%%%%%%%%%%%%%%%%%%%%%%%%%%%%%%%%%%%%%%%%%%%%%%%%%%%%%%%%%%%
%%%%%%%%%%%%%%%%%%%%%%%%%%%%%%%%%%%%%%%%%%%%%%%%%%%%%%%%%%%%%%%%%%%%%%%%%%%%%%%%%%%%%%%%%%%%%%%%
%%%%%%%%%%%%%%%%%%%%%%%%%%%%%%%%%%%%%%%%%%%%%%%%%%%%%%%%%%%%%%%%%%%%%%%%%%%%%%%%%%%%%%%%%%%%%%%%

\begin{figure}[!t]
\centering
\includegraphics[width=3.9in]{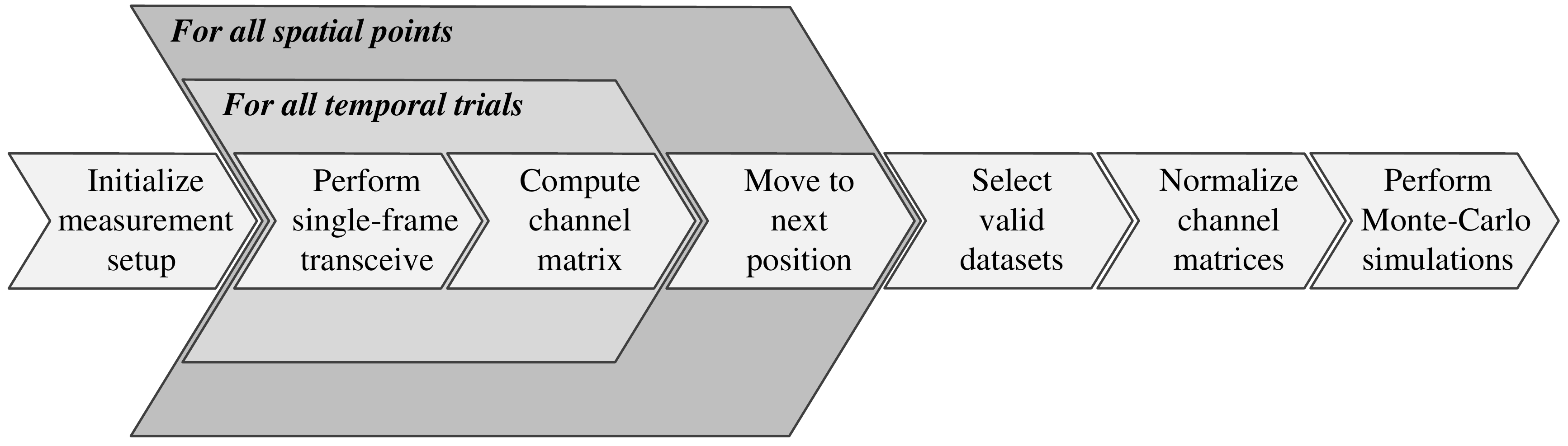}
\caption{Diagram outlining the experimental procedure for the measurement of the channel matrices and the computation of mutual information and error rate.}
\label{fig:diagram__}
\end{figure}

\textcolor{\markchanges}{
\section{Performance Analysis of Beam-Space MIMO} \label{sec_per_analy}
%In this section, the performance of the beam-space MIMO system is compared with that of conventional a $2{\times}2$ MIMO system under realistic propagation conditions. The section starts by outlining the experimental procedure and the performance criteria. Then, the measurement setup and scenarios are briefly described. Finally, the obtained results are presented and analyzed.  
Wireless systems are subject to strong multipath fading, as objects in the propagation environment tend to reflect, scatter, and diffract the propagating electromagnetic waves. Fading may lead to 20--30 dB fluctuations in the received signal strength when moving the transmit or receive antennas over very short distances, i.e., on the order of the wavelength. Therefore, taking measurements of a wireless system at only one transmitter-receiver position pair in space is insufficient for performance analysis. To achieve more representative performance results, it is necessary to spatially average by taking a high density of separate measurements for different spatial positions of the transmit (or receive) antenna(s) over an area of several wavelengths. Following the assumption of a stationary channel and reciprocity, results should be the same regardless of whether the transmit or receive antenna(s) are moved.
}
\textcolor{\markchanges}{
\subsection{Experimental Methodology and Procedure} \label{sec_per_proc}
Fig.~\ref{fig:diagram__} shows the experimental procedure adopted in this work for the performance analysis of beam-space MIMO (and conventional MIMO) in realistic environments. In the first phase, a frame-based measurement setup is employed to collect measurements of sufficient channel realizations across space. These channel measurements capture the combination of the propagation environment and the antenna radiation patterns, and thus allow the system performance to be evaluated without any assumptions of the statistics of the channel. At each spatial point, the measurement is repeated several times to reduce the impact of any temporal channel variations, for instance those caused by possible movement of people in the environment. Multiple temporal
measurements also allow any measurements corrupted by external interference in the same frequency band (e.g., 802.11 Wi-Fi transmissions) to be removed. For each spatial-temporal
channel realization, the channel matrix, denoted by $\mathbf{\widehat{H}}[s,t]$, where $s\in\{1,2,\ldots,\mathcal{S}\}$ and $t\in\{1,2,\ldots,\mathcal{T}\}$ are the discrete spatial and temporal indices, respectively (and $\mathcal{S}$ and $\mathcal{T}$ are the number of spatial points and temporal trials), is estimated by transmitting and receiving known training sequences as discussed earlier in Section~\ref{sec_ota_bsmimo}. The data sequences included in each frame are used to estimate the symbol error rates for measurement validation.
}
%the error rate corresponding to the test data sequences embedded in the transmitted frames is estimated, which will be used only for measurement validation purposes.

%%%%%%%%%%%%%%%%%%%%%%%%%%%%%%%%%%%%%%
\begin{figure}[!t]
\centering
\subfigure[]{\includegraphics[width=3.4in]{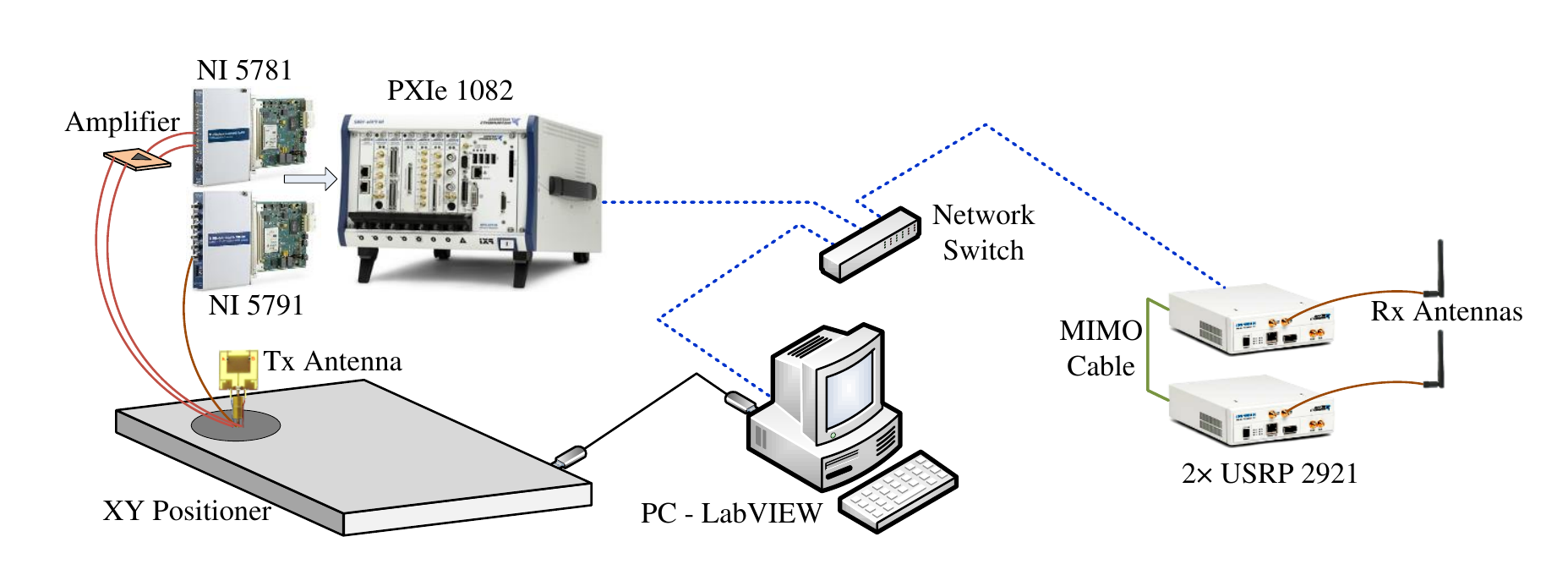} \label{fig:bs}}
\subfigure[]{\includegraphics[width=3.4in]{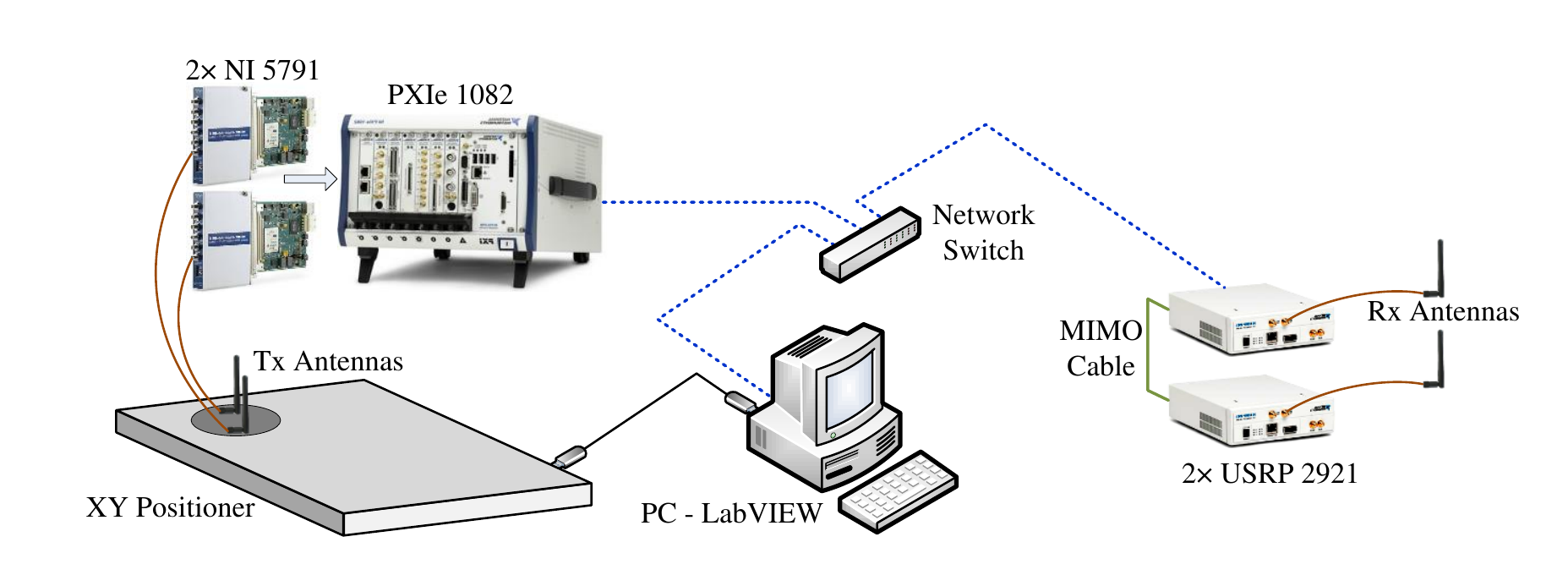} \label{fig:conv}}
\caption{(a) Experimental Platform used for the performance analysis of beam-space MIMO. (b) Experimental Platform used for the performance analysis of conventional 2$\times$2 MIMO.}
\label{fig:setup}
\end{figure}

%%%%%%%%%%%%%%%%%%%%%%%%%%%%%%%%%%%%%%%%%%
\textcolor{\markchanges}{
The link performance parameters are then extracted from the set of channel measurements. First, records with a measured error rate greater than a given threshold are removed from the data set. For each spatial measurement point, the average channel matrix, $\mathbf{\widehat{H}}_{\rm{av}}[s]$, is computed over valid temporal snapshots. The average channel matrices are then normalized to force unit average power at each receiver antenna, independent of the total power radiated from the transmitter, i.e., 
%%%%%
\begin{equation} \label{eq:normalization}
\left[\dfrac{1}{\mathcal{S}}\sum\limits_{s = 1}^{\mathcal{S}} {\left\|\mathbf{\widehat{H}}_{\rm{av,norm}}[s]\right\|^2}\right]^{1/2}=2 \,,
\end{equation}
%%%%%%%%%
where $\mathbf{\widehat{H}}_{\rm{av,norm}}[s]$ denotes the normalized channel matrices. 
The capacity can be estimated using the mutual information, which is computed by taking the expectation over the normalized channel matrices and noise using a Monte-Carlo technique~\cite{MI3,MI5}. Since the measured signal-to-noise ratio (SNR) from the experiments cannot be easily altered, a wide range of SNR values are considered by adding appropriately scaled independent and identically distributed (i.i.d.) Gaussian noise to the measurements. Similarly, the average symbol error rate (SER) is computed by running Monte-Carlo simulations over the measured channel matrices for different SNR operating points and independent of other factors such as transmit EVM and RF impairments. 
}
\textcolor{\markchanges}{
\subsection{Measurement Setup} \label{sec_OTA_measurementsetup}
Fig.~\ref{fig:bs} depicts the experimental platform used in our measurements of beam-space MIMO. This is very similar to the testbed described in Section~\ref{sec_testbedSetup}, and includes an XY positioner, allowing repositioning of the transmit antenna within an area of $310$~mm${\times}370$~mm (equivalent to $2.5\lambda_0{\times}3.0\lambda_0$ at 2.45 GHz, where $\lambda_0$ denotes the free-space wavelength) with a spatial accuracy of $0.1$~mm. 
}

\begin{figure*}
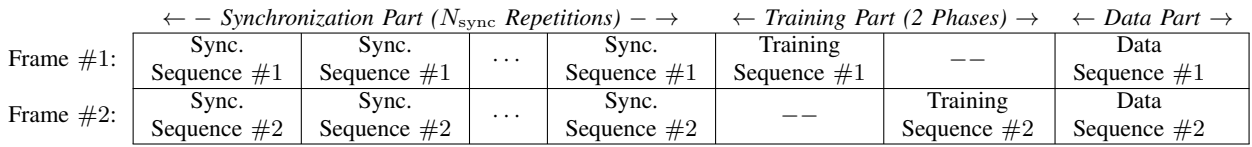

\centering
{
\footnotesize
{
\begin{tabular}{>{\raggedleft}m{1.5cm}|>{\centering}m{1.8cm}|>{\centering}m{1.8cm}|>{\centering}m{0.6cm}|>{\centering}m{1.8cm}|>{\centering}m{1.8cm}|>{\centering}m{1.8cm}|>{\centering}m{1.8cm}|}
\multicolumn{1}{c}{}    & \multicolumn{4}{c}{$\leftarrow -$ \emph{Synchronization Part (${N}_{\rm{sync}}$ Repetitions)} $-\rightarrow$ } & \multicolumn{2}{c}{$\leftarrow$  \emph{Training Part (2 Phases)} $\rightarrow$} & \multicolumn{1}{c}{$\leftarrow$  \emph{Data Part}  $\rightarrow$}\\
\cline{2-8}
Frame $\#1$: & Sync. Sequence $\#1$ & Sync. Sequence $\#1$ & $\cdots$ & Sync. Sequence $\#1$ & Training Sequence $\#1$ & $--$ & Data Sequence $\#1$     \\ 
\cline{2-8}
Frame $\#2$: & Sync. Sequence $\#2$ & Sync. Sequence $\#2$ & $\cdots$ & Sync. Sequence $\#2$ & $--$ & Training Sequence $\#2$ & Data Sequence $\#2$         \\ 
\cline{2-8}
\end{tabular} 
}
}
\caption{Frame structure used for conventional MIMO experiments. }
\label{fig:convmimo_frame1}
\end{figure*}

\textcolor{\markchanges}{
To perform measurements of conventional MIMO, the platform shown in Fig.~\ref{fig:conv} is used. The receiver side is the same as that of beam-space MIMO. However, at the transmitter side the NI~5781 (baseband) module is replaced with another NI~5791 (RF) module to produce the second RF signal. Moreover, two standard $2.4$~GHz antennas, identical to the ones at the receiving side, are utilized as transmit antennas. To carry out a set of measurements using either the beam-space MIMO platform or the conventional MIMO platform, the main LabVIEW program first initiates the positioner by moving the antenna(s) to the initial location. The program then calls the transmission and reception programs. 
%As expected, the transmission program for conventional MIMO measurements is different from the one discussed in Section~\ref{sec_ota_bsmimo} as the symbols of both frames are passed to their corresponding RF channels after oversampling and pulse-shaping. 
The frame structure of the conventional MIMO system differs slightly from that used in beam-space MIMO system, as shown in Fig.~\ref{fig:convmimo_frame1}. It can be seen that when each antenna transmits a synchronization or training sequence, the other antenna remains silent. 
%The system parameters used in our measurement campaign are selected according to Table~\ref{tab:parameters1}. We would like to stress that, although including the data segment in the frames is not essential in our experiments, we still dedicate half of the frame length to the data sequences. As stated in Section~\ref{sec_per_proc}, the information obtained from decoding the data segment is used to eliminate the measurements corrupted due to an incorrect frequency offset estimation or external interference.
According to the procedure outlined in Section~\ref{sec_per_proc}, once the measurements are repeated 11 times ($\mathcal{T}=11$) at a given spatial point, the main program moves the transmitting antennas to the next point. The same procedure is followed for all specified antenna locations. The measurements are taken over $400$ points ($\mathcal{S}=400$), and the spatial steps in the $x$ and $y$ directions are $16.3$~{mm} and $19.5$~mm, respectively. 
}

%%%%%%%%%%%%%%%%%%%%%%%%%%%%
%%%%%%%%%%%%%%%%%%%%%%%%%%%%%%%%%%

%%%%%%%%%%
%%%%%%%%%%%%
%%%%%%%%%%%%%%%%%%%%%%%%%%%%
%%\begin{figure}%[!b]
%%\centering
%%{\includegraphics[width=1.8in]{figures/spatial_points}} 
%%\caption{Measurement points for transmit antenna(s).}
%%\label{fig:spatial_points}
%%\end{figure}
%%%%%%%%%%%%%%%%%%%%%%%%%%%%%%%%%%

\section{Measurement Results} \label{sec_ota_results}
%%%
\begin{figure}[!t] 
\centering 
\subfigure[]{\includegraphics[width=2.2in]{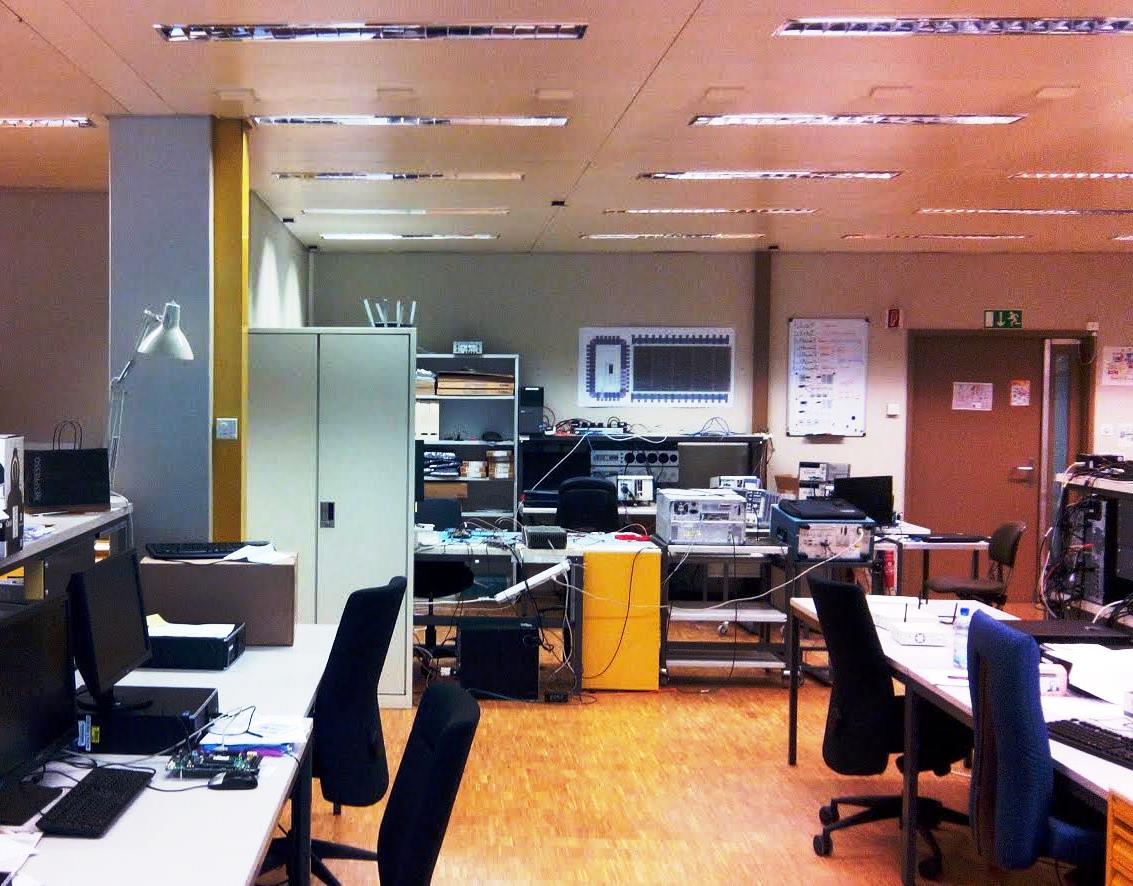} \label{fig:lab1}}
\subfigure[]{\includegraphics[width=2.6in]{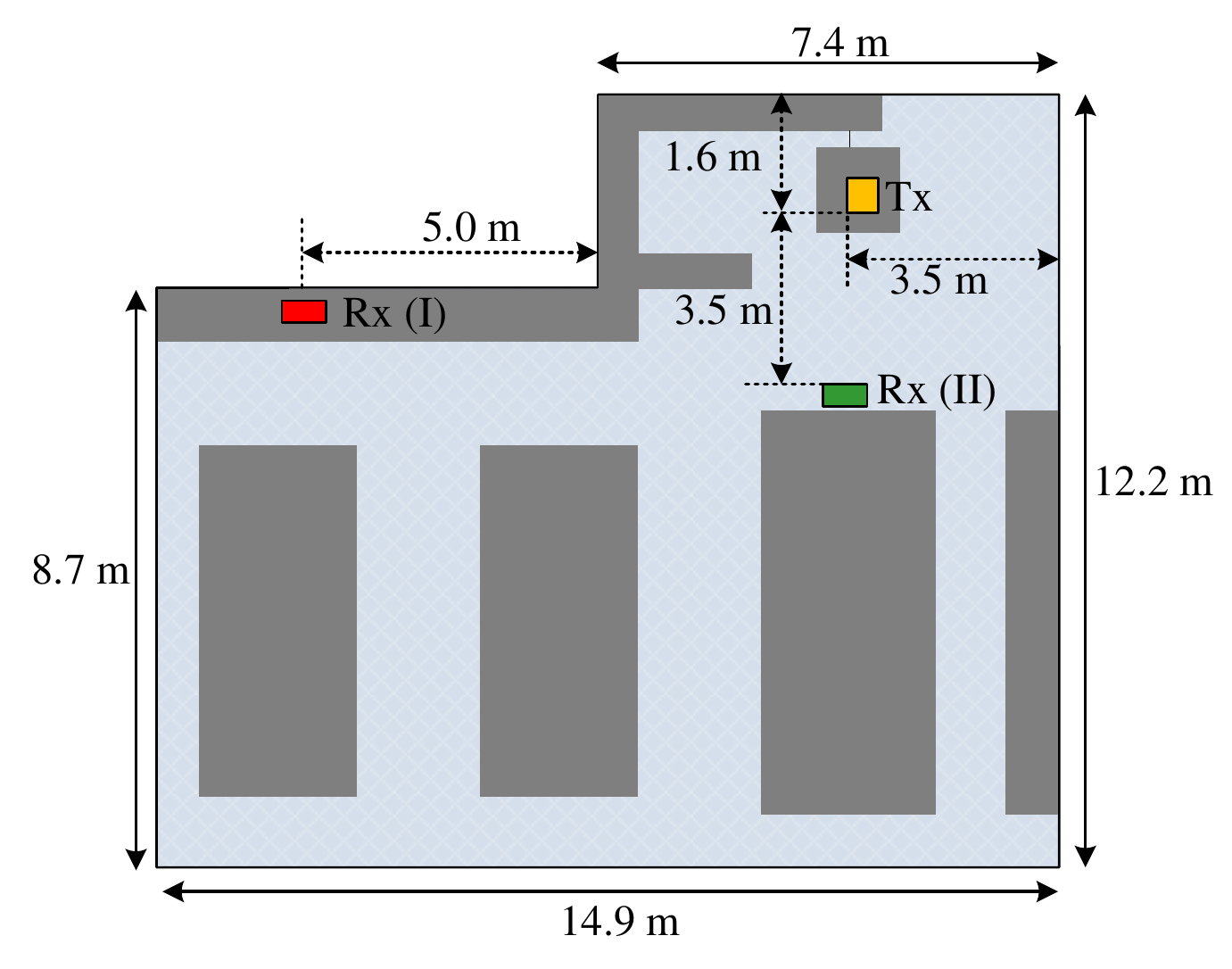} \label{fig:lab2}}
\caption{(a) Laboratory where the experiments were performed. (b) Floor plan of the measurement environment and the positions of the transmitter and the receiver. Grey areas represent the work desks and storage cabinets. The receiver location is marked by a red/green rectangle in the first/second scenario.}
\label{fig:lab_1_2} 
\end{figure}
%%%%%
The measurements were conducted in a laboratory environment at EPFL, shown in Fig.~\ref{fig:lab1}, with a considerable number of scattering objects, such as equipment, furniture, large metal storage cabinets, and other `clutter'. Fig.~\ref{fig:lab2} depicts the floor plan and the relative locations of the transmitter and receiver. Two different measurement scenarios were considered. In scenario I, there was no LOS path between the transmitter and the receiver. However, scenario II, was designed such that the channel became largely LOS, however, some NLOS components were also expected due to reflections from objects, walls, and ceiling. 

For each scenario, measurements were made for both the beam-space MIMO and conventional MIMO platforms. The experiments were conducted at night to reduce the effects of movement in the channel and unwanted external interference. Each set of $4400$ experimental measurements took approximately $7$ hours to complete. To minimize nonlinearities associated with the varactor diodes used as the variable loads in the beam-space MIMO antenna, the RF voltage swing on the diodes was kept much lower compared to their bias voltage. Accordingly, the RF power input to the beam-space MIMO antenna was limited to $0$~dBm. In conventional MIMO experiments, the RF power to each of the transmit antennas was $0$~dBm. This transmit power difference was later compensated when normalizing the channel matrices.

%\vspace{1mm}
%\noindent 
%\underline{\emph{Scenario I - NLOS}}

\subsection{Scenario I---NLOS}
\begin{figure}[!t] 
\centering 
\subfigure[]{\includegraphics[width=1.4in]{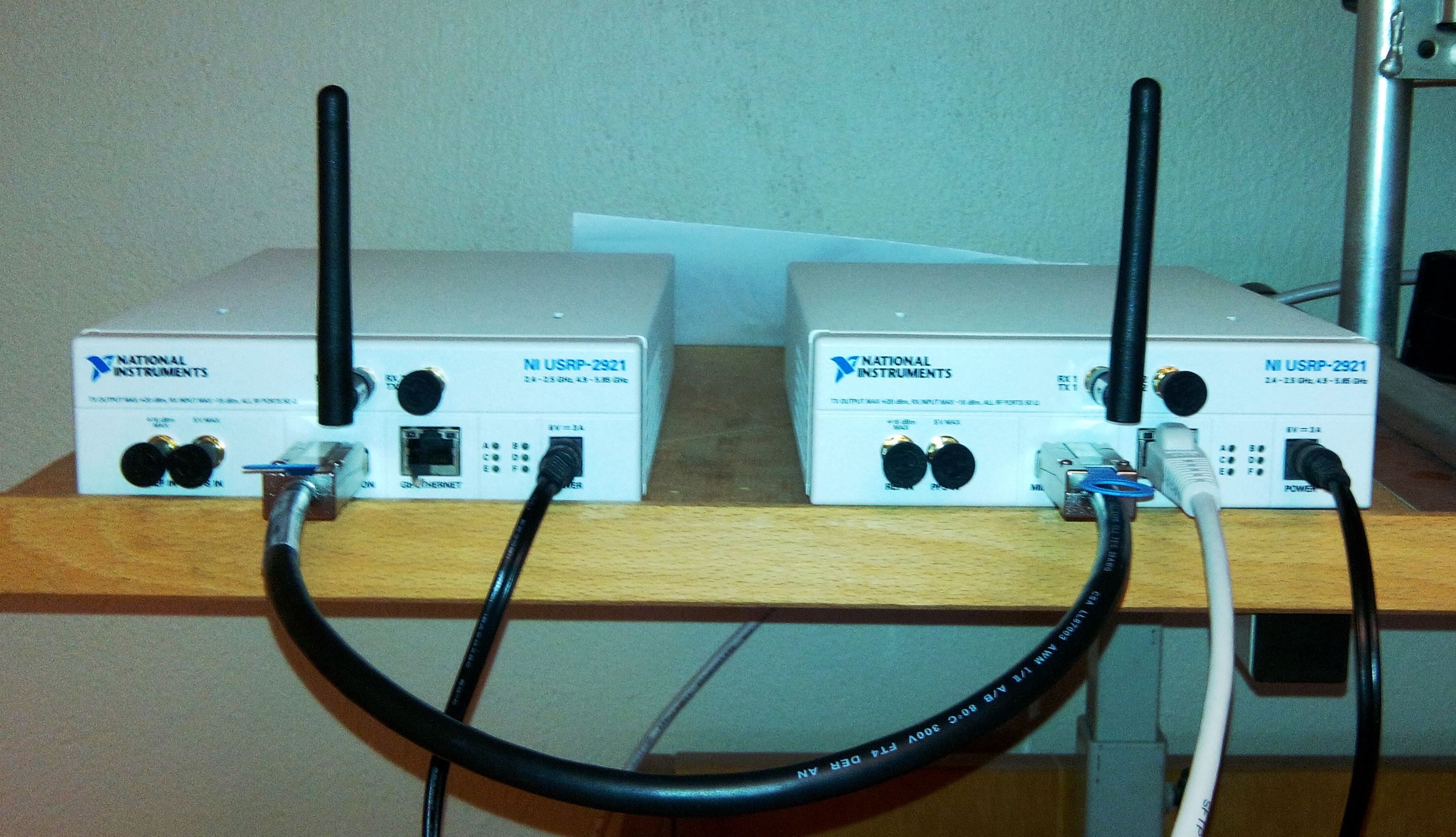} \label{fig:verpol}}  \quad \quad
\subfigure[]{\includegraphics[width=1.4in]{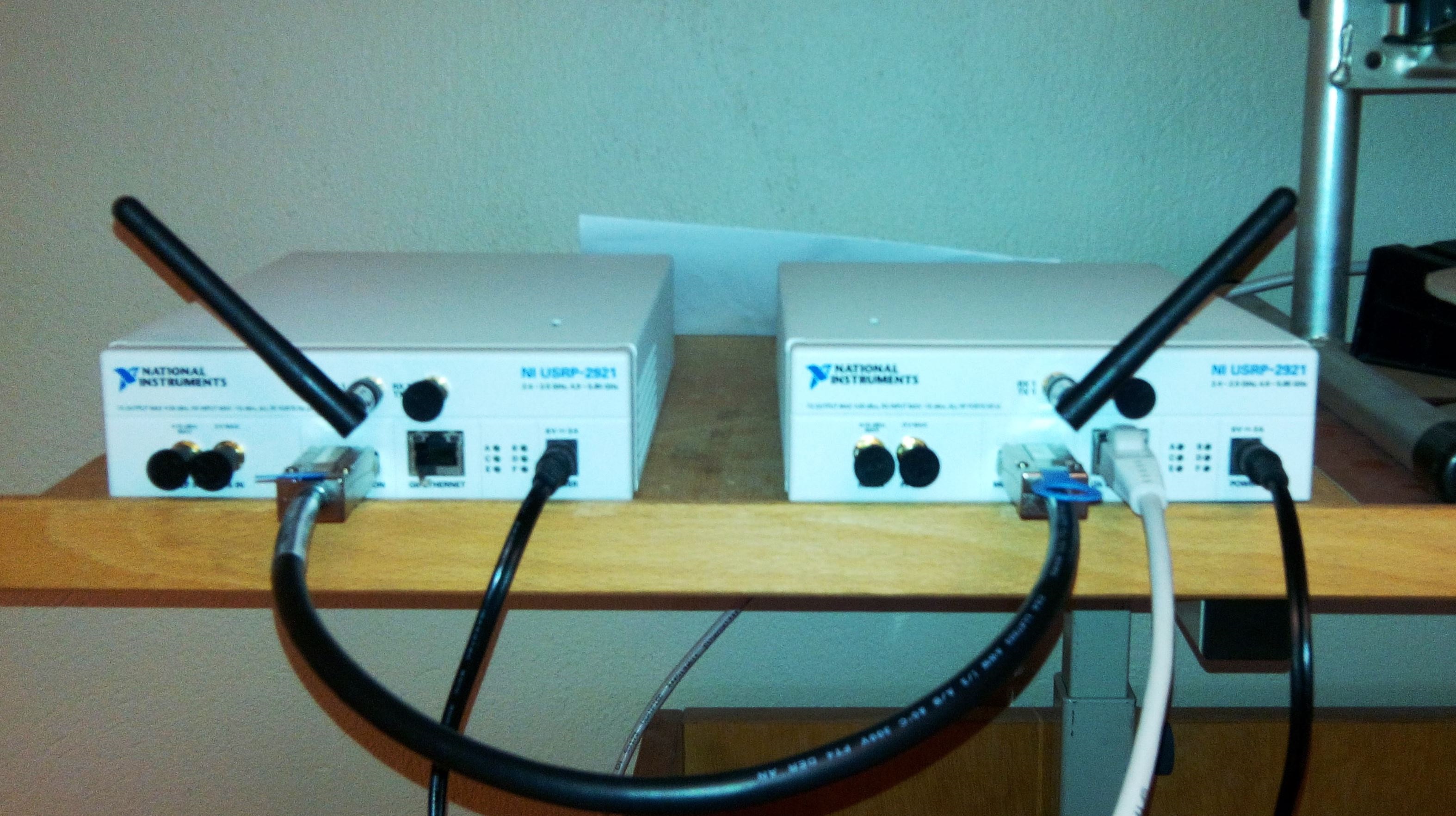} \label{fig:45pol}}
\caption{Polarization of the receive antennas; (a) vertically polarized, (b) perpendicular angled at $45^{\circ}$.}
\label{fig:Rxpol} 
\end{figure}

Three sets of experiments were conducted for this scenario:
%\vspace{-2mm}
\begin{itemize}
\item \emph{Experiment I.a:} The conventional MIMO platform was used, with all the transmit and receive antennas vertically polarized. The spacing between the transmit antennas was $120$~mm (approximately $\lambda_0$ at $2.45$~GHz), while the receive antennas were $240$~mm apart.
\item \emph{Experiment I.b:} The beam-space MIMO platform was used, with the receive antennas vertically polarized and $240$~mm apart as shown Fig.~\ref{fig:verpol}.
\item \emph{Experiment I.c:} The beam-space MIMO platform was used, with the receive antennas angled at $45^{\circ}$ and $240$~mm apart as shown in Fig.~\ref{fig:45pol}, to examine the impact of the relative polarization of the receive antennas.
\end{itemize}

Channel matrices for the measurements with an error rate greater than $10\%$ in the decoding of the data sequences in the same frame were excluded from the data set used in the analysis, as these were often found to be corrupted by external interference or had failed to correctly synchronize. %Table~\ref{tab:NLOSvalidity} shows the number of valid measurements recorded for each experiment. 
For all three experiments, in more than $88\%$ of the total $400$ spatial points there were at least one valid temporal measurement. Moreover, the measured SNR from the experiments was typically at least $20$~dB.

%%%%%%%%%%
%%%%%%%%%%%

\begin{figure}[!t]
  \subfigure[]{\label{fig:Iellip}  \includegraphics[width=3.2in]{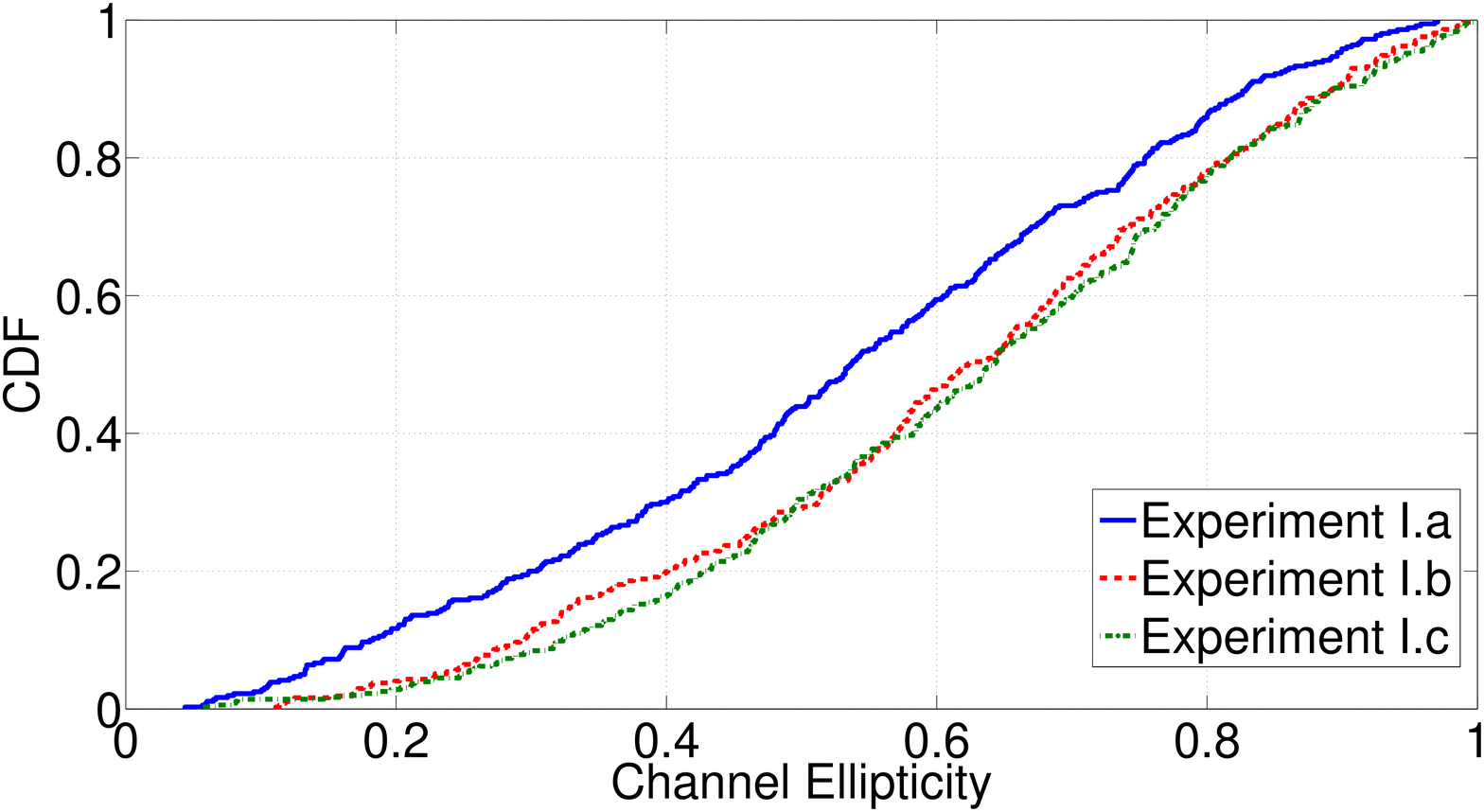}}
  \subfigure[]{\label{fig:Imi_ec}  \includegraphics[width=3.2in]{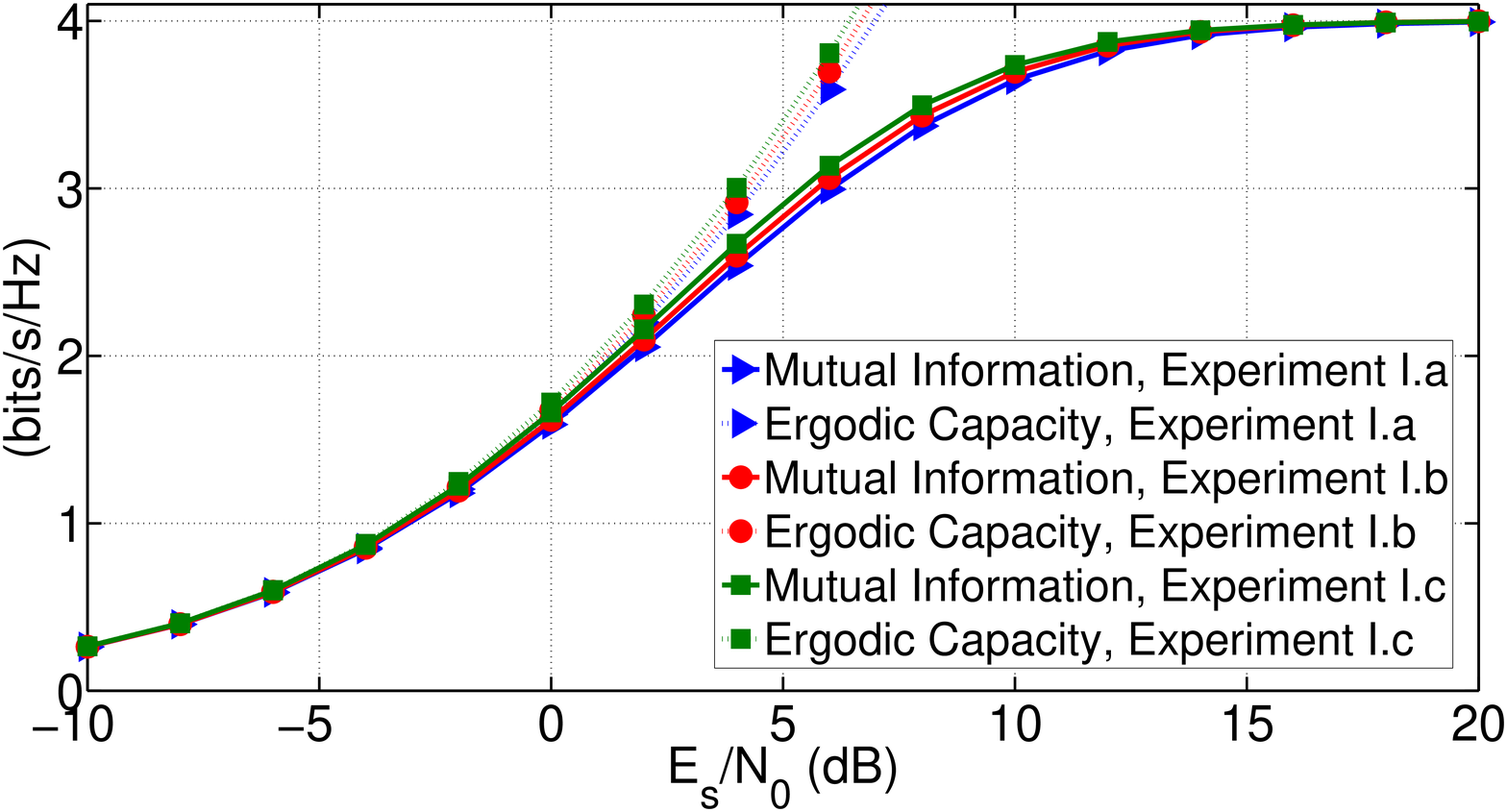}}
  \subfigure[]{\label{fig:Imi_cdf} \includegraphics[width=3.2in]{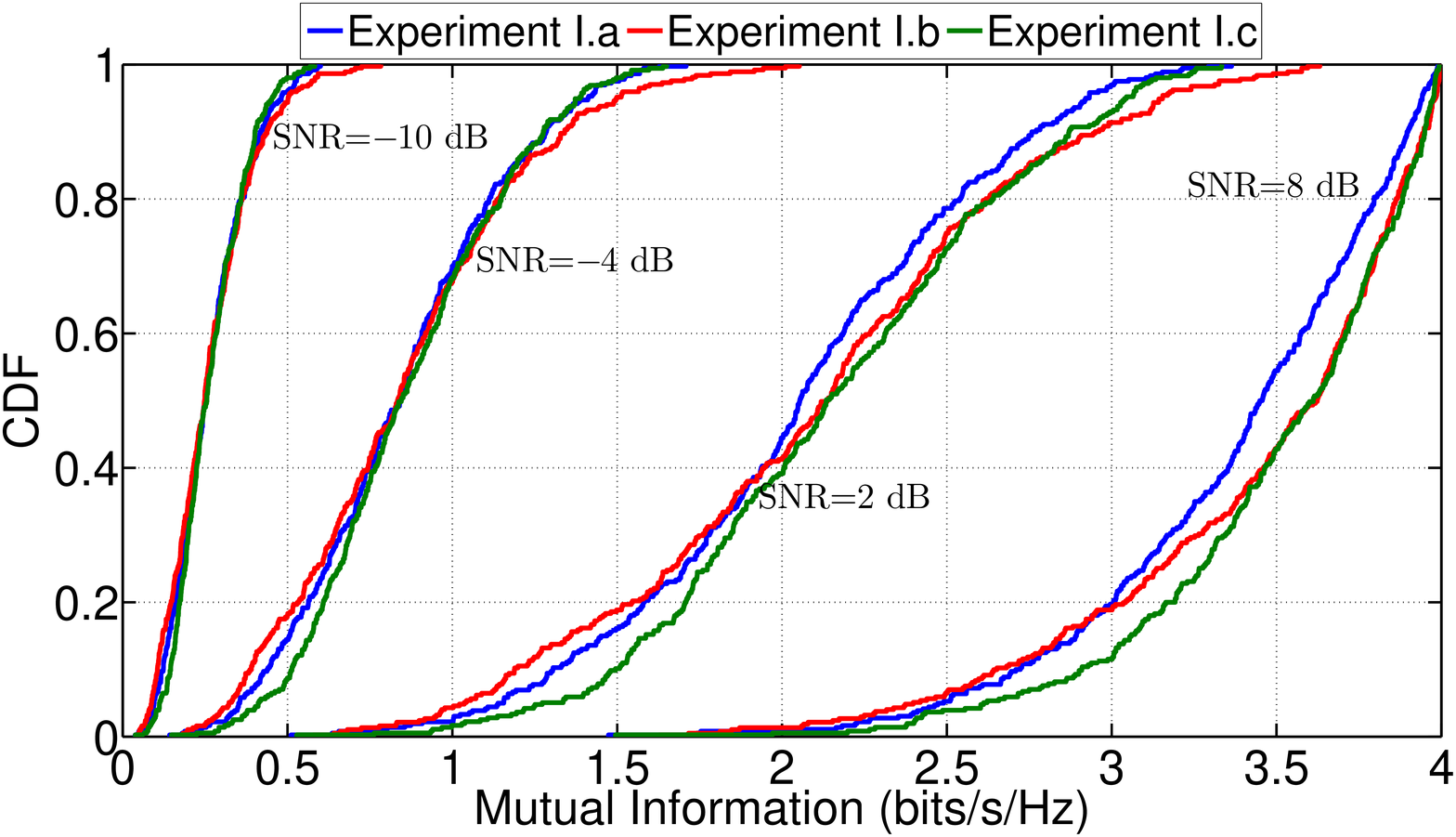}}
  \subfigure[]{\label{fig:I_SER}   \includegraphics[width=3.2in]{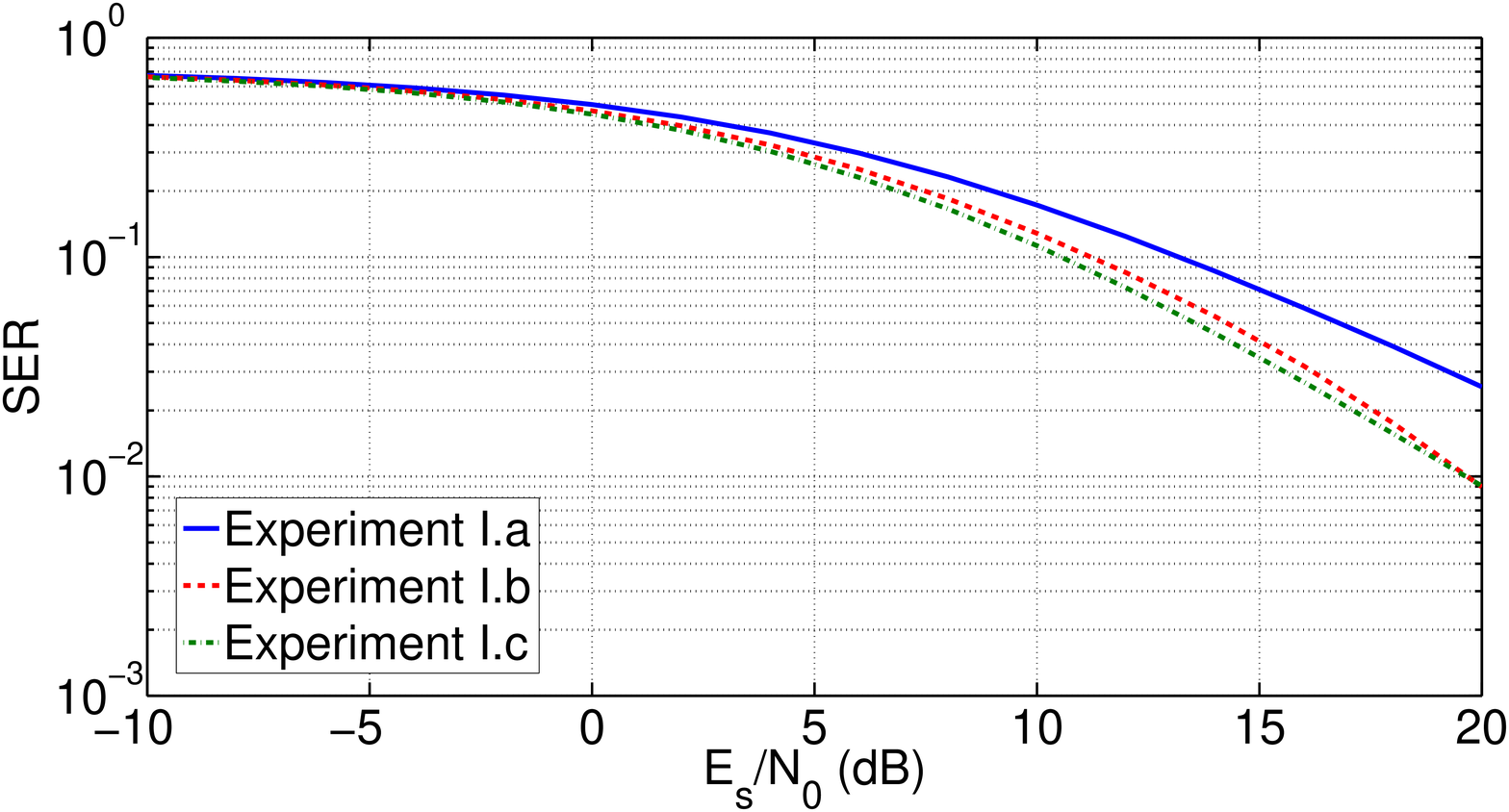}}
	\caption{Experimental measurement results for Scenario I---NLOS: (a) CDF of the channel ellipicity; (b) Mutual information and ergodic capacity curves; (c) CDF of the mutual information for different SNR values; and (d) SER obtained from Monte-Carlo simulations using the measured and normalized channel matrices.}
\end{figure}

%%%%%%%%%%%%%%%%%%%%%%%%%%%%%%%%%%%%
%%\begin{table}%[!b]
%%\caption{Valid Measurements in Scenario I}
%%\vspace{-2mm}
%%\label{tab:NLOSvalidity}
%%\renewcommand{\arraystretch}{0.5}
%%\centering
%%\footnotesize
%%\begin{tabular}{>{\centering}b{2.0cm}>{\centering}b{3.0cm}>{\centering}b{2.5cm}}
%%\toprule %\hline \hline \\[-1mm]
%% Set &  Total Valid Measurements  & Valid Spatial Points \\[1mm] %\hline \\
%% \midrule 
%%Experiment I.a & $3795$ ($86.3\%$) & $360$ ($90.0\%$) \\[1mm]%\hline \\
%% \midrule 
%%Experiment I.b & $3863$ ($87.8\%$) & $371$ ($92.7\%$) \\[1mm]%\hline \\
%% \midrule 
%%Experiment I.c & $3593$ ($81.7\%$) & $355$ ($88.7\%$) \\[1mm]
%% \bottomrule %\hline\hline 
%%\end{tabular}
%%\end{table}
%%%%%%%%%%%%%%%%%%%%%%%%%%%%%%%%%%%%%%%%%
Table~\ref{tab:IpowCOMP} summarizes the average received power for the three experiments, where the $3$~dB higher total input power in transmitting the data part in the conventional MIMO experiment was compensated to ensure a fair comparison against the beam-space MIMO experiments. It is observed that nearly the same (spatially averaged) power was received for the conventional and beam-space MIMO experiments. For each experiment, the difference between the average power at the two receiving antennas during the transmission of the data segment was less than $0.6$~dB. 
%%%%%%%%%%%%%%%%%%%%%%%%%%%%%%%%%%%%
\begin{table*}
\caption{Global Average of Average Received Power in Scenario I}
\vspace{-2mm}
\label{tab:IpowCOMP}
\renewcommand{\arraystretch}{0.5}
\centering
\footnotesize
\begin{tabular}{>{\centering}b{2.0cm}>{\centering}b{1.8cm}>{\centering}b{1.8cm}>{\centering}b{1.8cm}>{\centering}b{0.5cm}>{\centering}b{1.8cm}>{\centering}b{1.8cm}>{\centering}b{1.8cm}}
\toprule %\hline \hline \\[-1mm]
\multirow{2}{*}{set} & \multicolumn{3}{c}{{Average Received Power at Rx1 (dBm)}} & & \multicolumn{3}{c}{{Average Received Power at Rx2 (dBm)}} \\ [1mm]\cline{2-4} \cline{6-8} \\%[-1mm]
               &  Data &   1st Training &   2nd Training   &  & Data   &  1st Training &   2nd Training  \\[1mm] %\hline \\ 
\midrule
Experiment I.a & $-36.9$ & $-36.3$ & $-37.5$ & & $-36.4$ & $-35.4$ & $-37.7$  \\[1mm]%\hline \\
\midrule
Experiment I.b & $-36.0$ & $-37.1$ & $-35.1$ & & $-36.4$ & $-38.1$ & $-35.1$ \\[1mm]%\hline \\
\midrule
Experiment I.c & $-37.4$ & $-38.3$ & $-36.5$ & & $-36.8$ & $-36.5$ & $-36.8$ \\[1mm]
\bottomrule %\hline\hline 
\end{tabular}
\end{table*}
%%%%%%%%%%%%%%%%%%%%%%%%%%%%%%%%%%%%%%%%%
Table~\ref{tab:NLOSrce} gives the average received constellation error (RCE) of the data symbols at both antennas. For each experiment, the average RCE was calculated by averaging over all valid temporal snapshots and then by averaging over valid spatial points. It is observed that the average RCE of beam-space MIMO transmission is comparable with that of the conventional MIMO transmission.%, even with non-idealities in the beam-space MIMO antenna.
%%%%%%%%%%%%%%%%%%%%%%%%%%%%%%%%%%%%
\begin{table}[!t]
\caption{Average RCE of Received Data in Scenario I}
\vspace{-2mm}
\label{tab:NLOSrce}
\renewcommand{\arraystretch}{0.5}
\centering
\footnotesize
\begin{tabular}{>{\centering}b{2cm}>{\centering}b{2.6cm}>{\centering}b{2.6cm}}
\toprule %\hline \hline \\[-1mm]
 Set &  RCE (dB) at Rx1 & RCE (dB) at Rx2 \\[1mm] %\hline \\ 
\midrule
Experiment I.a & $-24.9$ & $-24.8$ \\[1mm]%\hline \\
\midrule
Experiment I.b & $-24.6$ & $-24.8$ \\[1mm]%\hline \\
\midrule
Experiment I.c & $-24.6$ & $-24.7$ \\[1mm]
\bottomrule
\end{tabular}
\end{table}
Using the normalized channel matrices, $\mathbf{\widehat{H}}_{\rm{av,norm}}[s]$, the full spatial correlation matrix can be calculated by~\cite{Oestges}
%%%%%%%%%%%%%%%%%%%%
\begin{equation*}
\boldsymbol{R}_{\mathbf{H}} := {\mathbb{E}}_{\mathbf{H}} \left[\rm{vec}(\mathbf{\widehat{H}}_{\rm{av,norm}})\rm{vec}(\mathbf{\widehat{H}}_{\rm{av,norm}})^{\rm{H}}\right] \,,
\end{equation*}
%%%%%%%%%%%%%%%%%%%%%%%%%%%%%%%%%%
where the $\rm{vec}(\cdot)$ operation stacks all columns of a matrix into a single column vector.
Table~\ref{tab:NLOS_R} shows the magnitude of $\boldsymbol{R}_{\mathbf{H}}$ for the three experiments. By comparing the diagonal entries, we find the power allocation in the beam-space MIMO channels is more uniform compared to the conventional MIMO channel. Moreover, the beam-space MIMO experiments yield slightly smaller correlation coefficients (in the off-diagonal entries of $\boldsymbol{R}_{\mathbf{H}}$). Table~\ref{tab:NLOS_R} also gives the average channel ellipticity for each experiment. The channel ellipticity, $\gamma[s] \in [0,1]$, is defined as the ratio of the geometric and arithmetic means of the channel eigenvalues for each spatial realization, and is used as a measure of multipath richness of a MIMO channel~\cite{ellipt}. Fig.~\ref{fig:Iellip} compares the cumulative distribution function (CDF) of the channel ellipticity in the three experiments. It is evident that the Eigenvalue spread of the beam-space MIMO channels is smaller.
%%%%%%%%%%%%%%%%%%%%%%%%%%%%%%%%%%%%
\begin{table}[!t]
\caption{Magnitude of Spatial Correlation Matrix and Channel Ellipticity in Scenario I}
\vspace{-2mm}
\label{tab:NLOS_R}
\renewcommand{\arraystretch}{0.5}
\centering
\footnotesize
\begin{tabular}{>{\centering}m{2cm}>{\centering}m{3.3cm}>{\centering}m{3.3cm}}
\toprule %\hline \hline \\[-1mm]
 Set &  $|\boldsymbol{R}_{\mathbf{H}}|$ & Channel Ellipticity \\%[1mm] \hline \\ 
\midrule
Experiment I.a & $\left[ {\begin{array}{*{20}{c}}
 1.09   & 0.96 &   0.36   & 0.25\\
    0.96  &  1.32 &   0.38 &    0.33\\
    0.36  &  0.38  &  0.82  &  0.22\\
    0.25  &  0.33  &  0.22 &   0.77
\end{array}} \right]$  & $0.53$ \\%[1mm]\hline \\
\midrule
Experiment I.b & $\left[ {\begin{array}{*{20}{c}}
 1.10   & 0.88 &   0.34   & 0.36\\
    0.88  &  1.03 &   0.29 &    0.37\\
    0.34  &  0.29  &  0.99  &  0.24\\
    0.36  &  0.37  &  0.24 &   0.88
\end{array}} \right]$ & $0.61$ \\%[1mm]\hline \\
\midrule
Experiment I.c & $\left[ {\begin{array}{*{20}{c}}
 0.97   & 0.84 &   0.21   & 0.25\\
    0.84  &  1.09 &   0.18 &    0.33\\
    0.21  &  0.18  &  0.88  &  0.13\\
    0.25  &  0.33  &  0.13 &   1.06
\end{array}} \right]$ & $0.62$ \\%[1mm]
\bottomrule %\hline\hline 
\end{tabular}
\end{table}
%%%%%%%%%%%%%%%%%%%%%%%%%%%%%%%%%%%%%%%%%

Fig.~\ref{fig:Imi_ec} compares the mutual information curves as a function of the SNR per symbol. While in the lower and higher SNR regimes all three experiments yield nearly identical mutual information values, the beam-space MIMO systems slightly outperforms in mid-range SNR values. In the conventional MIMO experiment, the power is equally distributed between the omnidirectional transmit antennas, which is not necessarily optimal for the specific NLOS environment under consideration. Fig.~\ref{fig:Imi_ec} also includes the curves of the ergodic capacity\footnote{Ignoring the finite constellation limitation.} as a validating tool for Monte-Carlo simulations. It is observed that in low-SNR regimes (up to $0$~dB) the effect of signaling using the finite QPSK alphabet is negligible.
Fig.~\ref{fig:Imi_cdf} depicts the empirical CDF of the mutual information values for each experiment.  While for low SNR values the conventional MIMO and beam-space MIMO experiments result in nearly identical distribution functions, in the mid-SNR region the beam-space MIMO systems outperforms the conventional MIMO system.
The symbol error rate curves obtained from Monte-Carlo simulations are shown in Fig.~\ref{fig:I_SER}. It is observed that the beam-space MIMO systems have better error performance compared to conventional MIMO.

%%%%%%%%%%%%%%%%%%55
%%%%%%%%%%%%%%%%%%%%%%%5

%%%%%%%%%%%%
%%\begin{figure}%%[!b] 
%%\centering 
%%\includegraphics[width=3.5in]{figures/ScenarioI_ellipticity} 
%%\caption{CDF of the channel ellipticity in Scenario~I.}
%%\label{fig:Iellip} 
%%\end{figure}
%%%%%%%%%%%%%
%%%%%%%%%%%%
%%\begin{figure}%%[!b] 
%%\centering 
%%\includegraphics[width=3.5in]{figures/ScenarioI_MI} 
%%\caption{Mutual information and ergodic capacity curves obtained from Monte-Carlo simulations using the measured and normalized channel matrices.}
%%\label{fig:Imi_ec} 
%%\end{figure}
%%%%%%%%%%%%%
%%%%%%%%%%%%
%%\begin{figure}%%[!b] 
%%\centering 
%%\includegraphics[width=3.5in]{figures/ScenarioI_MI_CDF} 
%%\caption{CDF of the mutual information for different SNR values.}
%%\label{fig:Imi_cdf} 
%%\end{figure}
%%%%%%%%%%%%%
%%%%%%%%%%%%
%%\begin{figure}%%[!b] 
%%\centering 
%%\includegraphics[width=3.5in]{figures/ScenarioI_SER} 
%%\caption{SER obtained from Monte-Carlo simulations using the measured and normalized channel matrices.}
%%\label{fig:I_SER} 
%%\end{figure}
%%%%%%%%%%%%%
%===================================================================================================================================================
%\vspace{1mm}
%\noindent 
%\underline{\emph{Scenario II---LOS}}
\subsection{Scenario II---LOS}

\begin{figure}[!t]
  \subfigure[]{\label{fig:IIellip}  \includegraphics[width=3.2in]{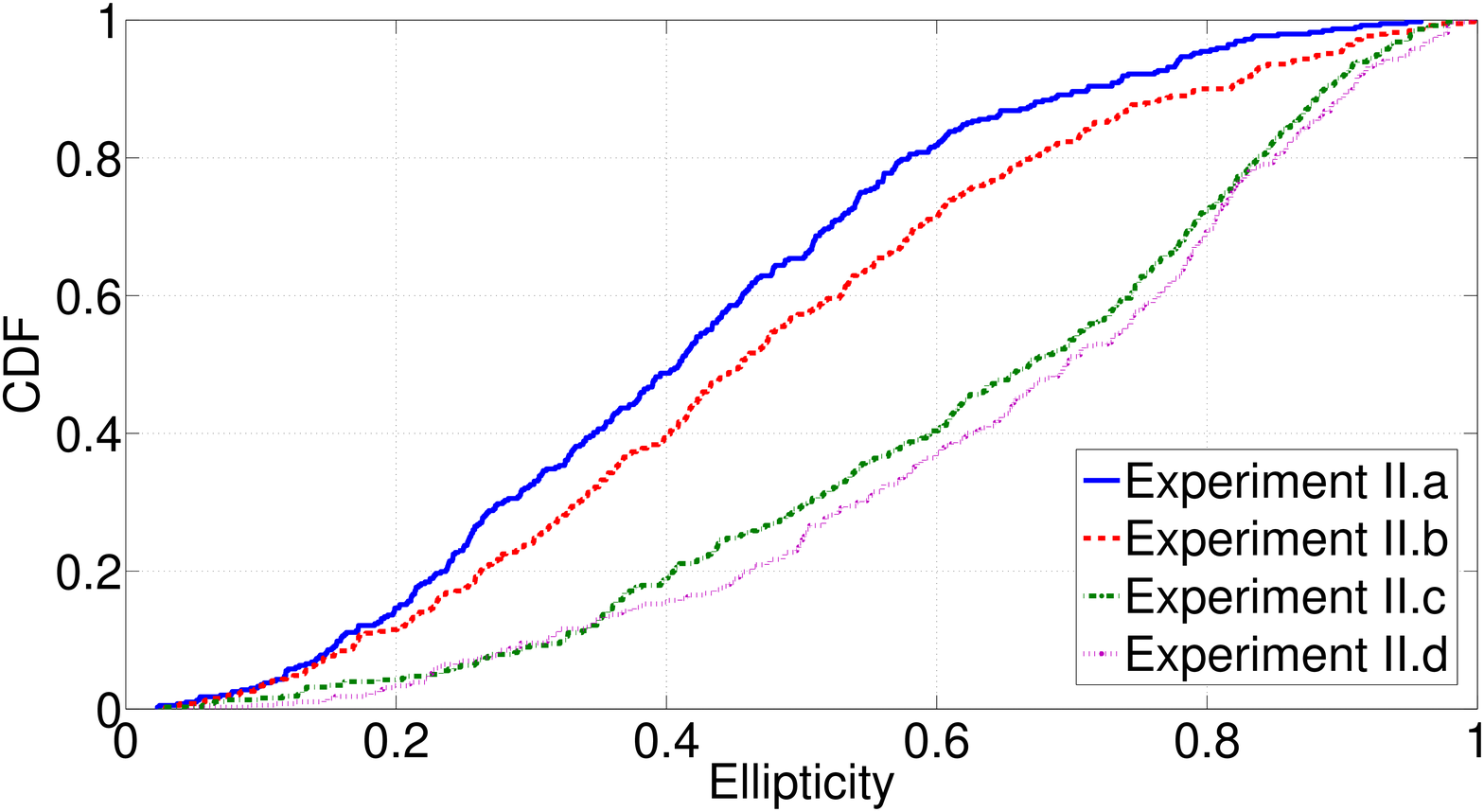}}
  \subfigure[]{\label{fig:IImi_ec}  \includegraphics[width=3.2in]{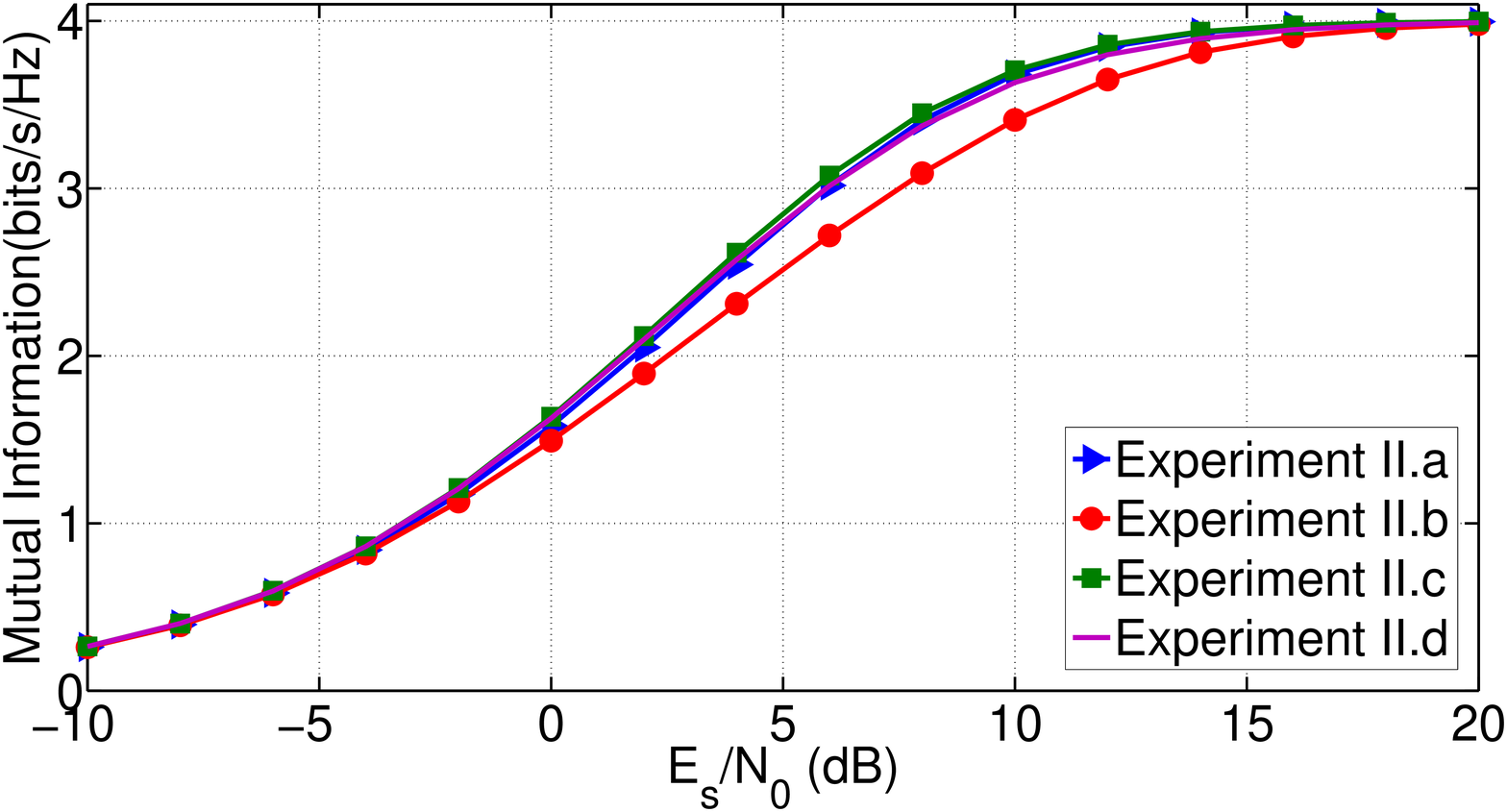}}
  \subfigure[]{\label{fig:IImi_cdf} \includegraphics[width=3.2in]{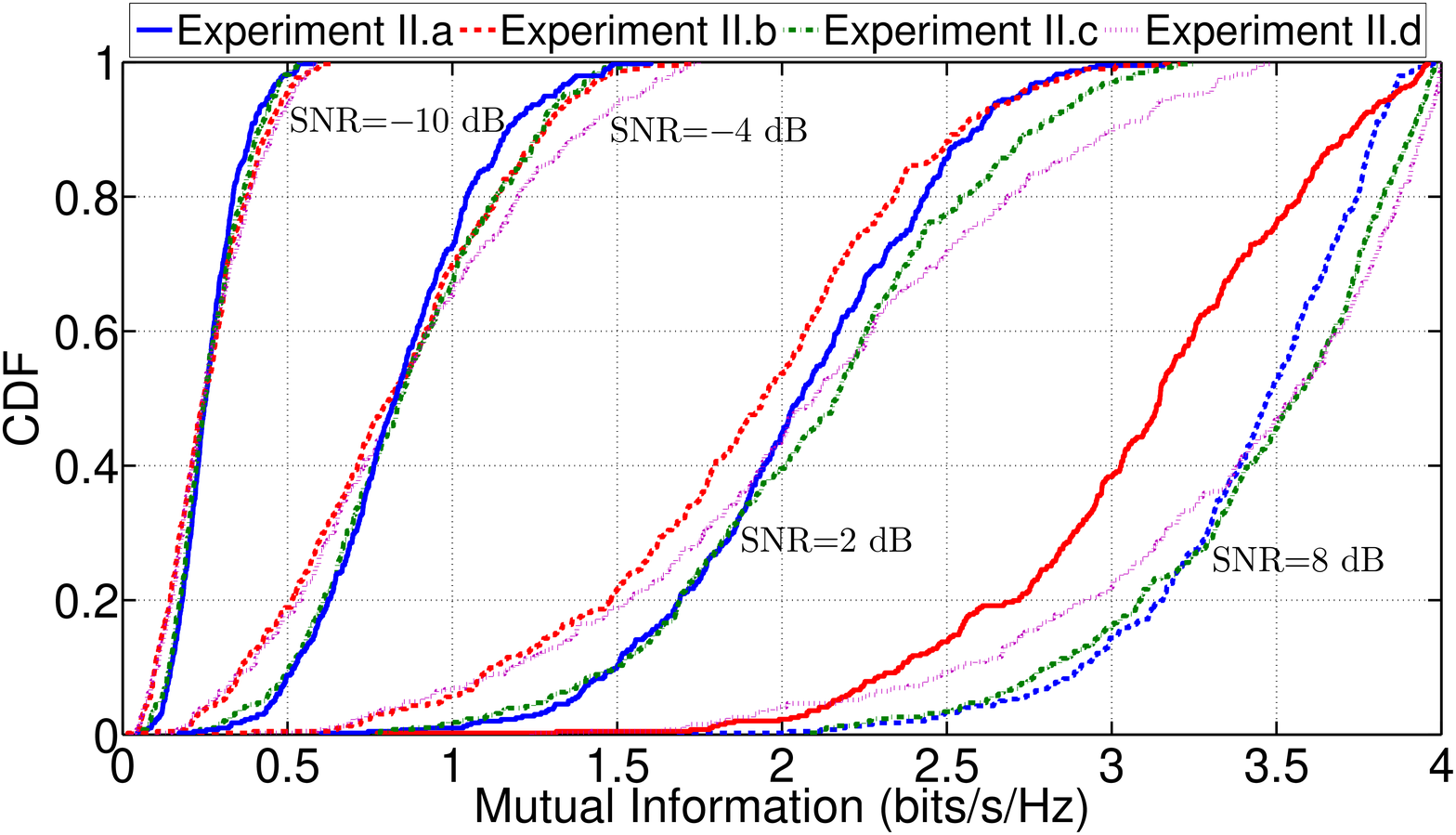}}
  \subfigure[]{\label{fig:II_SER}   \includegraphics[width=3.2in]{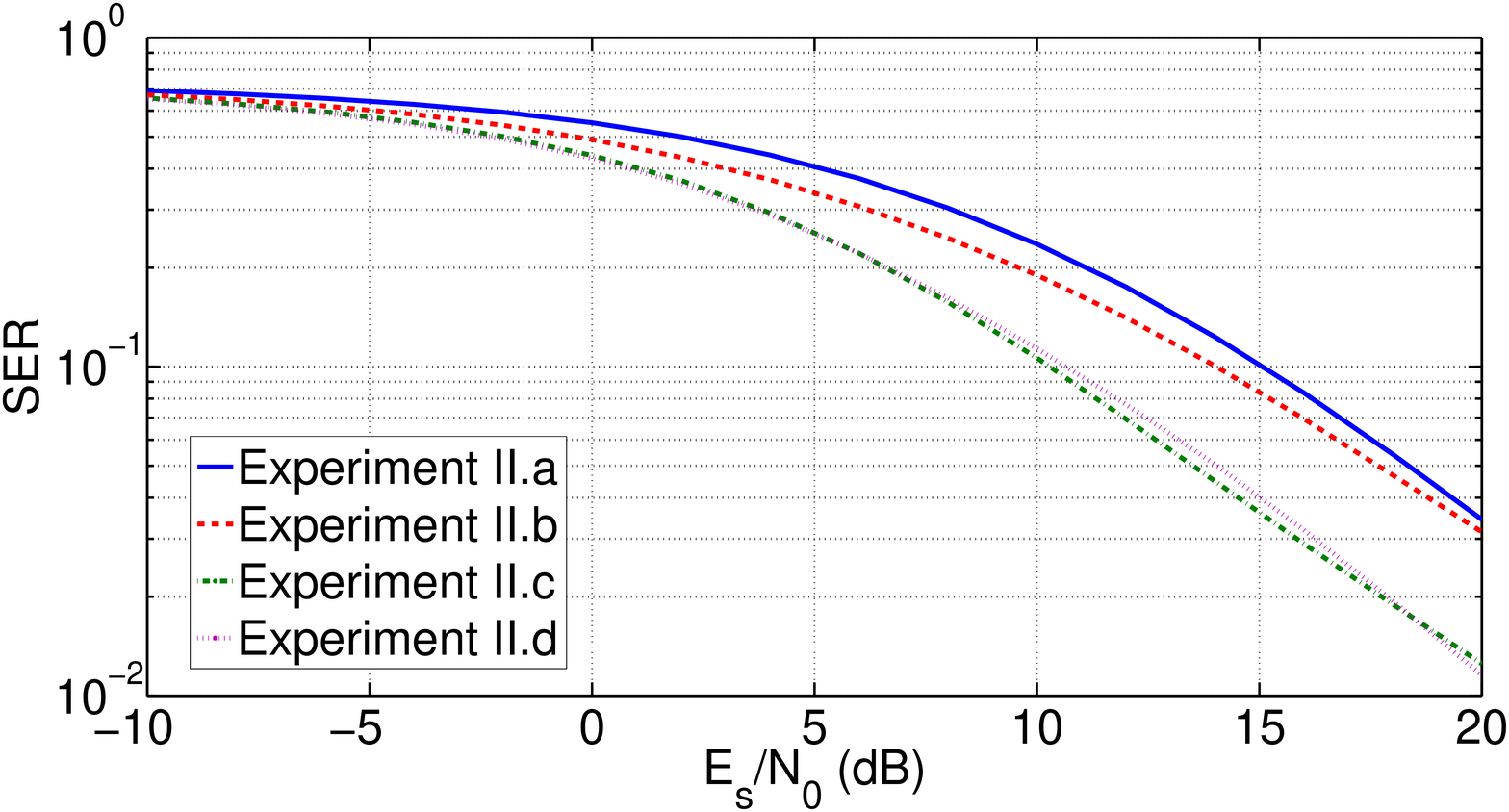}}
	\caption{Experimental measurement results for Scenario II---LOS: (a) CDF of the channel ellipicity; (b) Mutual information and ergodic capacity curves; (c) CDF of the mutual information for different SNR values; and (d) SER obtained from Monte-Carlo simulations using the measured and normalized channel matrices.}
\end{figure}

Four sets of experiments were conducted for this scenario:
%\vspace{-2mm}
\begin{itemize}
\item \emph{Experiment II.a:} The conventional MIMO platform was used, with all the transmit and receive antennas vertically polarized. The spacing between the transmit antennas was $120$~mm (approximately $\lambda_0$ at $2.45$~GHz), while the receive antennas were $240$~mm apart.
\item \emph{Experiment II.b:} The beam-space MIMO platform was used, with the receive antennas vertically polarized and $240$~mm apart as shown Fig.~\ref{fig:verpol}.
\item \emph{Experiment II.c:} The beam-space MIMO platform was used, with the receive antennas angled at $45^{\circ}$ and $240$~mm apart as shown in Fig.~\ref{fig:45pol}.
\item \emph{Experiment II.d:} The beam-space MIMO platform was used, with the receive antennas angled at $45^{\circ}$ and $240$~mm apart, while rotating the beam-space MIMO transmit antenna by $180^{\circ}$ about its vertical axis relative to the previous experiments.
\end{itemize}
%Table~\ref{tab:LOSvalidity} shows the number of valid measurements recorded for each experiment, using a $10\%$ measured error rate threshold.
For all four experiments, in more than $94\%$ of the total $400$ spatial points there were at least one valid temporal measurement (using a $10\%$ measured error rate threshold). Moreover, the measured SNR from the experiments was typically at least $25$~dB.
%%%%%%%%%%%%%%%%%%%%%%%%%%%%%%%%%%%%
%\begin{table}%[!b]
%\caption{Valid Measurements in Scenario II}
%\vspace{-2mm}
%\label{tab:LOSvalidity}
%\renewcommand{\arraystretch}{0.5}
%\centering
%\footnotesize
%\begin{tabular}{>{\centering}b{2.0cm}>{\centering}b{3.0cm}>{\centering}b{2.5cm}}
%\toprule %\hline \hline \\[-1mm]
% Set &  Total Valid Measurements  & Valid Spatial Points \\[1mm] %\hline \\ 
%\midrule
%Experiment II.a & $4265$ ($96.9\%$) & $396$ ($99.0\%$) \\[1mm] %\hline \\
%\midrule
%Experiment II.b & $4186$ ($95.1\%$) & $391$ ($97.7\%$) \\[1mm]%\hline \\
%\midrule
%Experiment II.c & $4017$ ($91.3\%$) & $379$ ($94.7\%$) \\[1mm]%\hline \\
%\midrule
%Experiment II.d & $4128$ ($93.8\%$) & $387$ ($96.7\%$) \\%[1mm]
%\bottomrule %\hline\hline 
%\end{tabular}
%\end{table}
%%%%%%%%%%%%%%%%%%%%%%%%%%%%%%%%%%%%%%%%%
On the assumption that the total input power to the transmit antenna(s) was $0$~dBm, Table~\ref{tab:IIpowCOMP} gives the global average of temporally-averaged received power for the four experiments. It is observed that the level of received power in the conventional MIMO experiment is, as expected, slightly higher than in the beam-space MIMO experiments. 
%%%%%%%%%%%%%%%%%%%%%%%%%%%%%%%%%%%%
\begin{table}[!b]
\caption{Global Average of Average Received Power in Scenario II}
\vspace{-2mm}
\label{tab:IIpowCOMP}
\renewcommand{\arraystretch}{0.5}
\centering
\footnotesize
\begin{tabular}{>{\centering}b{2.0cm}>{\centering}b{1.8cm}>{\centering}b{1.8cm}>{\centering}b{1.8cm}>{\centering}b{0.5cm}>{\centering}b{1.8cm}>{\centering}b{1.8cm}>{\centering}b{1.8cm}}
\toprule %\hline \hline \\[-1mm]
\multirow{2}{*}{set} & \multicolumn{3}{c}{{Average Received Power at Rx1 (dBm)}} & & \multicolumn{3}{c}{{Average Received Power at Rx2 (dBm)}} \\ [1mm] \cline{2-4} \cline{6-8} \\[-1mm]
               &  Data &   1st Training &   2nd Training   &  & Data   &  1st Training &   2nd Training  \\[1mm] %\hline \\ 
\midrule
Experiment II.a & $-23.6$ & $-24.7$ & $-22.8$ && $-22.8$ & $-23.5$ & $-22.2$  \\[1mm]%\hline \\
\midrule
Experiment II.b & $-23.5$ & $-23.7$ & $-23.3$ && $-24.3$ & $-23.4$ & $-25.5$ \\[1mm]%\hline \\
\midrule
Experiment II.c & $-23.2$ & $-22.6$ & $-23.8$ && $-23.6$ & $-23.6$ & $-23.5$ \\[1mm]%\hline \\
\midrule
Experiment II.d & $-24.5$ & $-26.5$ & $-23.0$ && $-23.6$ & $-22.5$ & $-25.0$ \\[1mm]
\bottomrule %\hline\hline 
\end{tabular}
\end{table}
%%%%%%%%%%%%%%%%%%%%%%%%%%%%%%%%%%%%%%%%%
The magnitude of $\boldsymbol{R}_{\mathbf{H}}$ and the average of the channel ellipticity are given in Table~\ref{tab:LOS_R}. The most unbalanced power allocation occurred in the Experiment II.b, while all off-diagonal correlation coefficients were relatively high in the conventional MIMO experiment. As shown in Fig.~\ref{fig:IIellip}, the beam-space MIMO experiments with angled receive antennas (Experiments II.c and II.d) yielded higher channel ellipticity.%, showing better performance in supporting parallel communication modes.
%%%%%%%%%%%%%%%%%%%%%%%%%%%%%%%%%%%%
\begin{table}[!t]
\caption{Magnitude of Spatial Correlation Matrix and Channel Ellipticity in Scenario II}
\vspace{-2mm}
\label{tab:LOS_R}
\renewcommand{\arraystretch}{0.5}
\centering
\footnotesize
\begin{tabular}{>{\centering}m{2cm}>{\centering}m{3.3cm}>{\centering}m{3.5cm}}
\toprule %\hline \hline \\[-1mm]
 Set &  $|\boldsymbol{R}_{\mathbf{H}}|$ & Channel Ellipticity \\[1mm] %\hline \\
 \midrule 
Experiment II.a & $\left[ {\begin{array}{*{20}{c}}
 0.70   & 0.71 &   0.57   & 0.25\\
    0.71  &  0.94 &   0.60 &    0.72\\
    0.57  &  0.60  &  1.10  &  0.66\\
    0.61  &  0.72  &  0.66 &   1.26
\end{array}} \right]$  & $0.42$ \\[1mm]%\hline \\
\midrule
Experiment II.b & $\left[ {\begin{array}{*{20}{c}}
 1.78   & 1.41 &   0.18   & 0.21\\
    1.41  &  1.54 &   0.15 &    0.20\\
    0.18  &  0.15  &  0.42  &  0.02\\
    0.21  &  0.20  &  0.02 &   0.26
\end{array}} \right]$ & $0.47$ \\[1mm]%\hline \\
\midrule
Experiment II.c & $\left[ {\begin{array}{*{20}{c}}
 1.32   & 1.02 &   0.19   & 0.07\\
    1.02  &  1.09 &   0.16 &    0.11\\
    0.19  &  0.16  &  0.77  &  0.31\\
    0.07  &  0.11  &  0.31 &   0.82
\end{array}} \right]$ & $0.63$ \\[1mm]%\hline \\
\midrule
Experiment II.d & $\left[ {\begin{array}{*{20}{c}}
 1.12   & 0.99 &   0.35   & 0.31\\
    0.99  &  1.24 &   0.33 &    0.35\\
    0.35  &  0.33 &  0.68  &  0.37\\
    0.31  &  0.35  &  0.37 &   0.95
\end{array}} \right]$ & $0.65$ \\[1mm]
\bottomrule %\hline\hline 
\end{tabular}
\end{table}
%%%%%%%%%%%%%%%%%%%%%%%%%%%%%%%%%%%%%%%%%
%%%%%%%%%%%%%

%%%%%%%%%%%%%%%%%%%%%%%%%%%%%%%%%%%%%%%%%%%%%%%%%%%%%%%%%%%%%%%%%%%%%%%%%%%%%%%%%%%%%%

The mutual information curves obtained from Monte-Carlo simulations are shown in Fig.~\ref{fig:IImi_ec}. The conventional MIMO system and two beam-space MIMO systems with angled receive antennas yield very similar capacity performance. The beam-space MIMO system with vertical receive antennas performed less well compared to the other three scenarios for mid-range SNR values, most likely due to poor power allocation. Fig.~\ref{fig:IImi_cdf} shows the empirical CDF of the mutual information values for each experiment. This result in particular indicates the performance of the beam-space MIMO system is dependent on the orientation of the transmit and receive antennas (this is hypothesized to occur as the basis patterns of the beam-space MIMO antenna are not omni-directional and similarly-polarized). Fig.~\ref{fig:II_SER} shows the SER curves obtained from Monte-Carlo simulations. It is observed that all three beam-space MIMO systems outperform the conventional MIMO system. However, the error performance depends on the relative polarization of the receive antennas.

%%%%%%%%%%%%
%%\begin{figure}%[b] 
%%\centering 
%%\includegraphics[width=3.5in]{figures/ScenarioII_ellipticity} 
%%\caption{CDF of the channel ellipticity in Scenario~II.}
%%\label{fig:IIellip} 
%%\end{figure}
%%%%%%%%%%%%%
%%%%%%%%%%%%
%%\begin{figure}%%[!b] 
%%\centering 
%%\includegraphics[width=3.5in]{figures/ScenarioII_MI} 
%%\caption{Mutual information and ergodic capacity curves obtained from Monte-Carlo simulations using the measured and normalized channel matrices.}
%%\label{fig:IImi_ec} 
%%\end{figure}
%%%%%%%%%%%%%
%%%%%%%%%%%%
%%\begin{figure}%%[!b] 
%%\centering 
%%\includegraphics[width=3.5in]{figures/ScenarioII_MI_CDF} 
%%\caption{CDF of the mutual information for different SNR values.}
%%\label{fig:IImi_cdf} 
%%\end{figure}
%%%%%%%%%%%%%
%%%%%%%%%%%%
%%\begin{figure}%[t] 
%%\centering 
%%\includegraphics[width=3.5in]{figures/ScenarioII_SER} 
%%\caption{SER obtained from Monte-Carlo simulations using the measured and normalized channel matrices.}
%%\label{fig:II_SER} 
%%\end{figure}

%%%%%%%%%%%%%%%%%%%%%%%%%%%%%%%%%%%%%%%%%%%%%%%%%%%%%%%%%%%%%%%%%%%%%%%%%%%%%%%%%%%%%%%%

\section{Conclusion}

The paper presented the design, realization, and measurement of the first reported QPSK beam-space MIMO antenna. Excellent agreement between the simulation and measurement results was obtained using an accurate design technique. The construction of an open-loop real-time beam-space MIMO testbed consisting of transmit and receive subsystems were described. \textcolor{\markchanges}{Accordingly, the first successful MIMO transmission of QPSK signals with a single RF chain was empirically demonstrated (employing the fabricated beam-space MIMO as the transmit antenna and a conventional two-element array as receive antennas). 
A measurement campaign was conducted to determine the performance of beam-space MIMO compared to conventional MIMO in real indoor channels. Experimental measurements of the channel matrix were carried out in LOS and NLOS scenarios for both MIMO systems. To correctly account for multipath fading, $400$ measurements were taken over a $310$~mm${\times}370$~mm area using an XY positioner. In addition, $11$ temporal measurements were made at each of the $400$ points to reduce the impact of movement in the channel and external interference. The mutual information and SER were computed by running Monte-Carlo simulations over the measured channel matrices and artificially generated i.i.d. Gaussian noise. Results for the NLOS scenario showed the mutual information performance of the beam-space and conventional MIMO systems were very similar, particularly at low- and high- SNR ranges. The beam-space MIMO system achieved up to $4.4\%$ higher mutual information than the conventional MIMO system for an SNR of $6$~dB. The SER performance comparison showed the beam-space MIMO system outperformed its conventional counterpart, resulting  in a gain of $3-3.5$~dB for an SER level of $0.03$. Both mutual information and SER in beam-space MIMO were slightly affected by the change in the polarization of the receive antennas. In particular, the LOS scenario confirmed that the directional-polarized basis patterns of the beam-space MIMO antenna could cause the system performance to depend on the orientation of receive and transmit antennas. 
While the mutual information of the beam-space MIMO system with vertically polarized receive antennas was approximately $10\%$ lower (at a SNR of $6$~dB) than the conventional MIMO system, the use of orthogonally-polarized receive antennas for beam-space MIMO regained the capacity performance losses.
Nevertheless, beam-space MIMO exhibited better SER performance compared to conventional MIMO for all NLOS and LOS scenarios considered.
}

\section*{Acknowledgment}
This work is dedicated to the memory of Professor Julien Perruisseau-Carrier. The authors would like to express gratitude to T. Debogovic, P. Belanovic, and R. Hochreutener at EPFL for their help in prototype fabrication, antenna characteristics measurements, and over-the-air experiments.  

\ifCLASSOPTIONcaptionsoff
  \newpage
\fi



%\bibliographystyle{IEEEtran}

%\begin{thebibliography}{1}


%\end{thebibliography}

\bibliographystyle{IEEEtran}
\bibliography{IEEEabrv,bibliography}


%\end{IEEEbiography}


%\newpage



%\vfill


\end{document}